\documentclass[journal=jctcce,manuscript=article,layout=twocolumn]{achemso}

\usepackage{color}
\usepackage{amsmath}
\usepackage[version=3]{mhchem} 

\def\molgw{\textsc{molgw}}
\def\br{\mathbf{r}}
\def\homo{\mbox{\scriptsize HOMO}}
\def\gga{\mbox{\scriptsize PBE}} 
\def\hf{\mbox{\scriptsize HF}}
\def\lr{\mbox{\scriptsize LR}}
\def\sr{\mbox{\scriptsize SR}}
\def\otrsh{\mbox{\scriptsize OTRSH}}

\graphicspath{{figures/}}

\author{Tonatiuh Rangel}
\affiliation{Molecular Foundry, Lawrence Berkeley National Laboratory, Berkeley, California 94720, United States}
\alsoaffiliation{Department of Physics, University of California, Berkeley, California 94720, United States}
\email{trangel@lbl.gov}

\author{Samia M. Hamed}
\affiliation{Molecular Foundry, Lawrence Berkeley National Laboratory, Berkeley, California 94720, United States}
\alsoaffiliation{Department of Physics, University of California, Berkeley, California 94720, United States}
\alsoaffiliation{Department of Chemistry, University of California, Berkeley, California 94720, United States}
\alsoaffiliation{Kavli Energy Nanosciences Institute at Berkeley, Berkeley, California 94720, United States}

\author{Fabien Bruneval}
\affiliation{Service de Recherches de M\'etallurgie Physique, CEA, Universit\'e Paris-Saclay, F-91191 Gif-sur-Yvette, France}
\alsoaffiliation{Molecular Foundry, Lawrence Berkeley National Laboratory, Berkeley, California 94720, United States}
\alsoaffiliation{Department of Physics, University of California, Berkeley, California 94720, United States}

\author{Jeffrey B. Neaton}
\affiliation{Molecular Foundry, Lawrence Berkeley National Laboratory, Berkeley, California 94720, United States}
\alsoaffiliation{Department of Physics, University of California, Berkeley, California 94720, United States}
\alsoaffiliation{Kavli Energy Nanosciences Institute at Berkeley, Berkeley, California 94720, United States}

\title{Evaluating the $GW$ approximation with CCSD(T) for charged excitations across the oligoacenes}


\begin{document}

\twocolumn[
\begin{@twocolumnfalse}
\begin{abstract}
Charged excitations of the oligoacene family of molecules, relevant for astrophysics and technological applications, are widely studied and therefore provide an excellent system for benchmarking theoretical methods.
In this work, we evaluate the performance of many-body perturbation theory~(MPBT) within the $GW$ approximation, relative to new high-quality CCSD(T) reference data, for charged excitations of the acenes.
We compare $GW$ calculations with a number of hybrid density functional theory starting points and with eigenvalue self-consistency.
Special focus is given to elucidating the trend of $GW$-predicted excitations with molecule length, from benzene to hexacene. 
We find that $GW$ calculations with starting points based on an optimally-tuned range separated hybrid (OTRSH) density functional and eigenvalue self-consistency can yield quantitative ionization potentials for the acenes. 
However, for the larger acenes, the predicted electron affinities can deviate considerably from reference values. 
Our work paves the way for predictive and cost-effective $GW$ calculations of charged-excitations of molecules and identifies certain limitations of current $GW$ methods used in practice for larger molecules.
\end{abstract}
\end{@twocolumnfalse}
]

\section{Introduction}

\begin{figure}[h!]
\includegraphics[width=3.3 in]{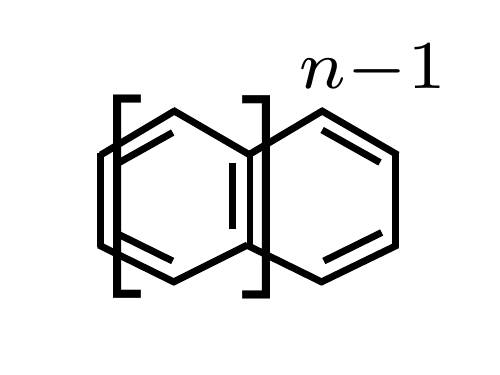}
\caption{The acenes general formula.}
\label{fig:acenes}
\end{figure}

The oligoacene molecules belong to a class of aromatic hydrocarbons consisting of linearly fused benzene rings (Fig.~\ref{fig:acenes}).
This family of molecules has been studied in the context of variety of opto-electronic applications, and in particular, the larger acenes and their derivatives are used for field-effect transistors\cite{anthony_functionalized_2006} and in solar-cell devices.\cite{forrest_introduction:_2007,bredas_molecular_2009,smith_singlet_2010,lee_singlet_2013}
In addition, acenes and other polycyclic aromatic hydrocarbons are abundant in the universe and their properties are of importance to astrophysics.\cite{_space_,mulas_estimated_2006,boersma_nasa_2014}

Charged excitations, namely excited states associated with electron addition and removal, have been well-studied for acene molecules with a variety of computational approaches, including density functional theory~(DFT), many-body perturbation theory~(MBPT), and wavefunction-based quantum chemistry methods,\cite{hajgato_benchmark_2008,deleuze_benchmark_2003,dupuy_vertical_2015} and hence constitute an excellent benchmark case for the development and refinement of methods.
Additionally, aspects of charged excitations for acenes are still not entirely understood. 
For example, anion states of benzene and naphthalene are unbound, and hence challenging to measure;\cite{burrow_electron_1987} further, whether measured excitations are strictly “vertical” or “adiabatic” can be ambiguous. \cite{hajgato_benchmark_2008,deleuze_benchmark_2003,tamblyn_simultaneous_2014}

Numerous theoretical approaches can be used to compute charged excitations, including wavefunction-based methods, such as coupled cluster techniques and quantum Monte Carlo~(QMC), as well as DFT and MBPT.
While wavefunction-based methods are regarded as highly accurate and the “gold standard”, they exhibit poor scaling and are currently intractable for many complex systems. 
On the other hand, MBPT within the $GW$ approximation,\cite{hedin_new_1965} a Green's function-based approach built upon Kohn-Sham DFT wavefunctions and eigenvalues, scales more modestly with the number of basis functions and is broadly applicable to a range of molecules, solids, surfaces, and nanostructures. 
It is therefore useful to quantify the performance of $GW$ approaches relative to high-accuracy wavefunction-based methods for molecular systems, such as the acenes.

Previous studies\cite{bruneval_$gw$_2009,x._blase_first-principles_2011,marom_benchmark_2012,korzdorfer_strategy_2012,gallandi_long-range_2015,knight_accurate_2016,rangel_structural_2016} have benchmarked $GW$ calculations against experiment and couple-cluster techniques for small- to medium-sized molecules, including some acenes.
In particular, these works have examined the performance of different DFT starting points and self-consistent $GW$ (see Section.~\ref{sect:mbpt-theory} for details), and have found that some $GW$ approaches are more predictive for charged excitations in organic molecules than others. 
However, as valuable as these studies are, they, by design, were not all-inclusive; further, there has yet to be a report of trends for charged excitations across a series of acenes of increasing size.

In this work, we calculate ionization potentials (IPs) and electron affinities (EAs) of acenes, from benzene to hexacene, using MBPT within the $GW$ approximation. 
We compare our $GW$ results to highly-accurate coupled-cluster calculations, with single, double, and perturbative triple excitations [CCSD(T)]. 
Since prior CCSD(T) reference calculations relied on extensive extrapolations, especially for the larger acenes, we perform new CCSD(T) calculations and include a comparison to these new results.
For completeness, we also benchmark recently-developed exchange-correlation DFT functionals that make use of a system-dependent non-empirically determined amount of exact exchange via the optimally-tuned range-separated hybrid~(OTRSH) class of functionals.\cite{kronik_excitation_2012}
Special attention is given to the accuracy of approximations within $GW$: we test convergence issues; the performance of DFT starting points for $GW$, including global hybrids and OTRSHs; and the performance of eigenvalue self-consistent $GW$ approaches (see Section~\ref{sect:mbpt-theory} for details).

\section{Theoretical methods}

For small gas-phase molecules, such as the acenes considered here, in principle, IPs and EAs can be determined from DFT via total energy differences between charged and neutral species.\cite{ziegler_approximate_1991}
However, this $\Delta$SCF approach is limited to frontier orbital energies, ill-defined for states  above the vacuum energy~(i.e., with negative EA), and can be inaccurate especially for large molecules due to the nature of approximate DFT exchange-correlation functionals.\cite{cohen_insights_2008,vlcek_deviations_2015,gallandi_accurate_2016} 
In this work, we produce a quantitative benchmark of ab initio MBPT within the $GW$ approximation,  an alternative and more general approach for electron  addition and removal energies of acene molecules.
Our $GW$ calculations are based on DFT and, as we will show below, they are quantitatively dependent on the solutions to the underlying generalized Kohn-Sham equations (and therefore sensitive to the functional used).
Moreover, recent developments in generalized Kohn-Sham DFT suggest that appropriately-constructed exchange-correlation functionals can lead to accurate charged-excitation spectra.\cite{refaely-abramson_fundamental_2011,korzdorfer_long-range_2011,kronik_excitation_2012,refaely-abramson_quasiparticle_2012,refaely-abramson_gap_2013,egger_outer-valence_2014,gallandi_accurate_2016}
In the following sub-sections, we describe our $GW$ approach after first summarizing a class of range-separated hybrid ~(RSH) functionals that we use both as a starting point for our $GW$ calculations and as an independent reference for comparison with $GW$.

\subsection{Optimally-tuned range separated hybrid functionals}

We first consider two representative tuned range-separated hybrid DFT schemes: the Baer-Neuhauser-Lifshitz~(BNL) functional\cite{baer_density_2005,livshits_well-tempered_2007} (see details in Ref.~\citenum{kuritz_charge-transfer-like_2011}) and a recent OTRSH functional,\cite{kronik_excitation_2012,refaely-abramson_quasiparticle_2012,refaely-abramson_gap_2013} described below.
OTRSH partitions the Coulomb operator to balance exact exchange with generalized gradient approximation~(GGA) exchange and correlation as\cite{leininger_combining_1997,yanai_new_2004}
\begin{equation}
\frac{1}{r_{12}} = \frac{\alpha + \beta \textrm{erf}(\gamma r_{12})}{r_{12}} + \frac{1 - (\alpha + \beta \textrm{erf}(\gamma r_{12}))}{r_{12}},
\label{eq.range-separation}
\end{equation}
where, in this case, the first term is treated with Hartree Fock~(HF) and the second is treated with the GGA Perdew-Burke-Ernzerhof~(PBE)  functional.\cite{perdew_generalized_1996}
This partition leads to the following form of the OTRSH exchange correlation energy:
\begin{eqnarray}
 E^{\otrsh}_{xc} &=&  
  \alpha E_{x}^{\hf}  
  + ( 1 -\alpha - \beta ) E_{x}^{\gga}
  \nonumber\\ 
  &+&\beta E_{x}^{\hf,\lr}(\gamma)
  + \beta E_{x}^{\gga,\sr}(\gamma)
  + E_{c}^{\gga},\nonumber\\
\end{eqnarray}
where $E_{c}^{\gga}$ is the PBE correlation energy and the mixing parameters $\alpha$, $\beta$, and  $\gamma$ are tuned so as to fulfill exact conditions and theorems of DFT.\cite{kronik_excitation_2012}

Although $\alpha$ can be determined from first principles in some cases,\cite{srebro_does_2012,refaely-abramson_quasiparticle_2012,luftner_experimental_2014,egger_outer-valence_2014} we follow Refs.~\citenum{refaely-abramson_quasiparticle_2012},~\citenum{refaely-abramson_gap_2013} and \citenum{rohrdanz_long-range-corrected_2009} and set $\alpha$ to 0.2, corresponding to a fraction of short-range exchange similar to that of a conventional hybrid functional.
Additionally, we set $\alpha + \beta = 1$ to enforce long-range asymptotic exact exchange.
Then, $\gamma$, the range-separation parameter, is varied to achieve a minimization of the target function
\begin{eqnarray}
  J^2(\gamma) &=& \left[\mathrm{IP}^\gamma(N) + E^\gamma_{\homo}(N)\right]^2 \nonumber\\
  &+& 
  \left[\mathrm{IP}^\gamma(N+1) + E^\gamma_{\homo}(N+1)\right]^2,
\end{eqnarray}
where $\mathrm{IP}^{\gamma}(N)$ is determined via a $\Delta SCF$ approach from total energy differences as $\mathrm{IP}^{\gamma}(N)=\epsilon^{\gamma}_\textrm{tot}(N-1)-\epsilon^{\gamma}_\textrm{tot}(N)$, where $\epsilon^{\gamma}_{\textrm{tot}}(N)$ and $\epsilon^{\gamma}_{\textrm{tot}}(N-1)$ are total energies of the neutral and cation species respectively.
This procedure is often referred to as ``gap-tuning'',\cite{gallandi_accurate_2016} and in the limit of vanishing $J^2(\gamma)$, this enforces the ionization potential theorem of DFT, namely that minus the energy of the Kohn-Sham highest occupied molecular orbital~(HOMO) equals the first ionization potential energy.\cite{perdew_density-functional_1982,salzner_koopmans_2009,levy_exact_1984,perdew_comment_1997,almbladh_exact_1985}  
For both benzene and naphthalene, the $N+1$ anionic state is unbound, so only the first of these two terms is minimized. 
Within this framework, the optimal $\gamma$ parameters for benzene through hexacene are found to be 0.25, 0.21, 0.19, 0.17, 0.15, and 0.14~bohr$^{-1}$ respectively.

In this work, all OTRSH and BNL calculations are performed with the \textrm{Q-Chem 4.2} software package,\cite{shao_advances_2015} and all geometries are relaxed with \textrm{Q-chem} with DFT using the B3LYP\cite{becke_new_1993,lee_development_1988} functional and a cc-pVTZ basis set.

\subsection{Many body perturbation theory within the $GW$ approximation}
\label{sect:mbpt-theory}

In MBPT, the $GW$ approximation consists of a closed set of equations for the Green's function $G$, the screened-exchange $W$, and the electronic self-energy
\begin{multline}
  \Sigma(\br,\br',\omega)=i \int d\omega'
   e^{i\delta\omega'} G(\br,\br',\omega+\omega') \\
    \times
     W(\br',\br,\omega') ,
\end{multline}
which is non-local, non-Hermitian, and frequency dependent.\cite{hedin_new_1965,chelikowsky_quantum_1996,aryasetiawan_gw_1998,onida_electronic_2002,bruneval_quasiparticle_2014}
A fully self-consistent solution of the $GW$ equations using large basis sets is currently unfeasible for most systems of interest and further approximations are required.
Most frequently, the $GW$ self-energy is applied perturbatively as a first-order correction to generalized Kohn-Sham states obtained from a DFT calculation.
This is the so called ``one-shot'' $GW$ or $G_0W_0$, in which for a given $i^\textrm{th}$ state, the Kohn-Sham~(KS) wavefunction $|i\rangle$ is kept constant and the corresponding eigenvalue $E_i$ is corrected, as follows:\cite{hybertsen_electron_1986}
\begin{equation}
E^{\textrm{QP}}_i= E^\textrm{KS}_i+ \langle i | \Sigma(\omega=E^{\textrm{QP}}_i) - v_{xc}| i \rangle .
\label{eq:gw}
\end{equation}
It follows directly from the previous expression that the one-shot $GW$ result may depend much on the quality of the underlying exchange-correlation~(XC) functional.

A well-known workaround is to use an XC functional whose generalized Kohn-Sham mean-field spectrum is closer to the actual charged excitation energies, a so-called improved starting point, e.g. hybrid functionals.\cite{bruneval_benchmarking_2013,korzdorfer_strategy_2012,atalla_hybrid_2013,korbel_benchmark_2014,gallandi_long-range_2015,govoni_large_2015,knight_accurate_2016}
In this work, we will consider some promising hybrid functionals that have been identified in previous studies,\cite{bruneval_benchmarking_2013,bruneval_systematic_2015} including PBE0\cite{perdew_rationale_1996} and BHLYP,\cite{becke_new_1993} with 25\% and 50\% exact exchange, respectively.
We will also consider the OTRSH functional\cite{livshits_well-tempered_2007} described above.
In the following, we indicate the mean-field starting point with an ``@'' sign, i.e., $G_0W_0$@PBE0,  $G_0W_0$@BHLYP and $G_0W_0$@OTRSH, to refer to one-shot $GW$ on top of PBE0, BHLYP and OTRSH, respectively.

Another approach to mitigate starting-point dependence would be to perform a self-consistent calculation.\cite{stan_fully_2006,rostgaard_fully_2010,caruso_unified_2012}
An approximate self-consistent scheme that only updates the eigenvalues entering $\Sigma$, while keeping the KS wavefunctions frozen, has been highlighted for molecules recently\cite{x._blase_first-principles_2011,jacquemin_benchmarking_2015} with promising results.
This scheme is known as eigenvalue self-consistent $GW$, or ``ev$GW$'', and involves the iterative updating of the eigenvalues in both the Green's function $G$ and the screened Coulomb interaction $W$.
A partial eigenvalue self-consistent scheme that only updates eigenvalues in $G$ and not in $W$ has been proposed and used extensively for solids.\cite{shishkin_kresse_2007}
This approach, ``ev$GW_0$'', has been previously reported to be less effective for molecules,\cite{jacquemin_benchmarking_2015} but we test it here for completeness.
In this work, we perform three to four iterations to converge the partial self-consistent $GW$ results within 0.01~eV; when using a ``good'' $GW$ starting point (with energies close to the ev$GW$ solution, such as BHLYP) only two iterations are needed to reach the same convergence threshold for the molecules considered here.

\subsection{$GW$ calculations in a Gaussian basis}
Our $GW$ calculations, and the computations generating our DFT starting points, are performed with \molgw~\cite{molgw_cite} code, using Gaussian basis sets.
A comprehensive description of this code can be found in Refs.~\citenum{bruneval_systematic_2015},~\citenum{bruneval_benchmarking_2013} and \citenum{bruneval_ionization_2012}; but briefly, after a self-consistent DFT calculation, \molgw{} evaluates the $GW$ self-energy via a spectral representation of the dynamical polarizability~$\chi$, allowing analytical calculation of the self-energy without any loss of information; i.e., $\chi$ is calculated exactly in a given basis set.
\molgw{} makes use of external libraries for the evaluation of electron repulsion integrals, \textsc{libint}, \cite{libint2} and for exchange-correlation potentials, \textsc{libxc}.\cite{marques_libxc:_2012}

The present study uses relatively large basis sets; e.g., hexacene C$_{26}$H$_{16}$ in aug-cc-pVTZ requires as many as 1564 basis functions.
To deal with these large systems, four-center integrals are evaluated approximately via an approach referred to as the resolution-of-the-identity in the Coulomb metric\cite{vahtras_cpl1993,weigend_pccp2002}.
This approximation has been used successfully in past $GW$ calculations,\cite{rohlfing_prb1995,x._blase_first-principles_2011,ren_njp2012} and it leads to a drastic reduction in the computational burden: the scaling of the atomic to molecular orbital transforms is reduced to $N^4$ from $N^5$.
More specifically, this method involves approximating the 4-center electron repulsion integrals
according to
\begin{equation}
\label{eq:ri1}
  ( \alpha \beta | \frac{1}{r} | \gamma \delta) \approx
    \sum_{PQ} ( \alpha \beta | \frac{1}{r} | P )   
     ( P |\frac{1}{r} | Q )^{-1}  ( Q | \frac{1}{r} | \gamma \delta)  ,
\end{equation}
where M\"ulliken notation is used. 
The Greek letters represent the basis functions for the wavefunction,
whereas the capitals $P$ and $Q$ run over an auxiliary basis set.
In practice, we use an approach\cite{molgw1} in which the square root of the matrix $( P |\frac{1}{r} | Q )$ is calculated
and thus the evaluation of Eq.~(\ref{eq:ri1}) is further accelerated.

The accuracy of the approximation in Eq.~(\ref{eq:ri1}) relies critically on the ability of the auxiliary basis set to represent the Coulomb interaction properly.
In this work, we use the well-established auxiliary basis sets of Weigend,\cite{weigend_pccp2002} an atom-centered basis consistent with the Dunning basis.\cite{dunning_jcp1989}
We have explicitly determined that approximate use of the resolution-of-the-identity affects the $GW$ energies by at most 1~meV in the case of benzene.

\textsc{molgw} analytically treats the frequency dependence of $\Sigma$ 
by calculating the polarizability $\chi$ within the random-phase approximation.
Then $\chi$ is written as a matrix containing all the single excitations available in the basis set, except for the carbon 1$s$ states which are kept frozen.
Therefore, the only convergence criteria is the basis set size, which will be carefully checked below.

\subsection{Basis set convergence}
\begin{figure*}[h!]
\begin{tabular}{@{\hskip 0 in}c@{\hskip 0.01 in}c@{\hskip 0.01 in}c}
\includegraphics{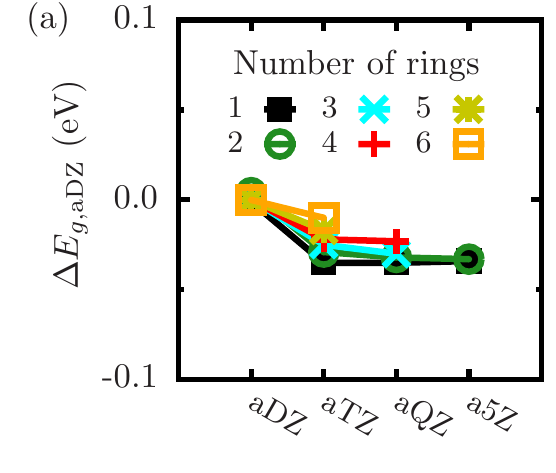}&
\includegraphics{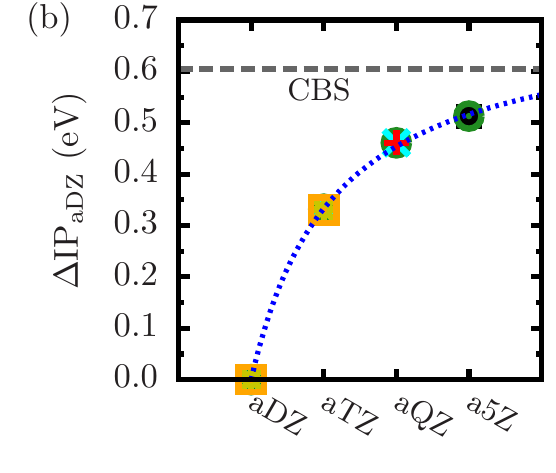}&
\includegraphics{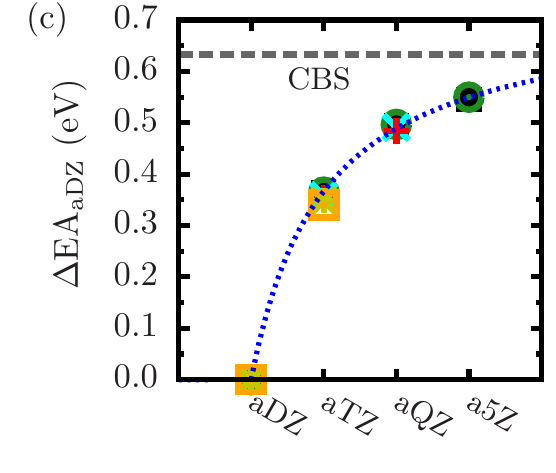}
\end{tabular}
\caption{\small Convergence of charged excitations with respect to the basis set size for the oligoacene molecules within $GW$ based on a PBE0 starting point. 
Calculated IP--EA gap energies~($E_g$) converge fast with respect to the basis set size, whereas IPs and EAs are extrapolated to the basis set~(CBS) limit using a function of the form $f(x)=a+b/(x-x_0)$ (dotted-blue lines).
Note that as the molecule size increase, calculations with large basis sets become unfeasible, and hence some points are omitted in the figures.
For convenience, we show the energy difference with respect to the results obtained with the \text{aDZ} basis, $\Delta{E}_{\mbox{\scriptsize aDZ}}=E-E_{\mbox{\scriptsize aDZ}}$;
In fact, points for different molecules overlap making evident that all quantities converge at similar rates for the different molecules considered here.
}
\label{fig:convergence}
\end{figure*}

As mentioned above, in this work we use the augmented basis sets of Dunning,\cite{dunning_jcp1989} which are designed to converge smoothly towards the complete basis set limit (CBS).
For simplicity, we refer to these basis sets as aDZ, aTZ, and so on, instead of their full-length names aug-cc-pVDZ, aug-cc-VTZ, etc.

In Figure~\ref{fig:convergence}, we show the convergence of our calculated values for the ionization potential~(IP), the electron affinity~(EA), and the IP--EA gap with respect to basis set size for benzene to hexacene.
All results are obtained with $G_0W_0$@PBE0.
The basis set is increased in size in the aug$-$pVnZ series from n$=$D to 5.
To better compare the rate of convergence across the acene series, we have set their aDZ values equal to zero in each case.
For anthracene ($n=3$) and larger acenes, we forgo some calculations with the largest basis sets, aug$-$pV5Z and aug$-$pVQZ, due to the significant computational burden.

We find that the calculated IPs and EAs converge monotonically and are well-fit with a simple function, as indicated in Figure~\ref{fig:convergence}.
Remarkably, the IPs and EAs of all acene molecules converge in the same manner, independent of the length of the molecule.
In fact, the energy difference between IPs/EAs calculated with the aTZ basis set and the CBS limit is $\sim$0.26~eV.
Thus, one may evaluate IPs and EAs at the CBS limit for these molecules by first performing $GW$ calculations with the aTZ basis set, and then adding $0.26$~eV.

\section{Ionization potentials and electron affinities}

\subsection{Obtaining reference values}
\label{sect:ccsdt}

Here, we revisit CCSD(T) IP energies in the CBS limit for acene molecules, following the focal point analysis~(FPA) approach laid out in Ref.~\citenum{deleuze_benchmark_2003}.
In the FPA, CCSD(T) best estimates are obtained from single-point calculations at the restricted-HF level and adding incremental improvements to the correlation energy at the second-, third- and partial fourth-order M{\o}ller-Plesset levels (MP2, MP3 and MP4SDQ). 
In turn, these are followed by improvements from coupled-cluster calculations including singles and doubles~(CCSD) and a perturbative estimate of the triples~(CCSD(T)), and by extrapolating to the CBS limit using Dunning basis sets of increasing size.
Interestingly, the data in Ref.~\citenum{deleuze_benchmark_2003} exhibit a significant break of $\sim$0.1~eV in the trend along the oligoacene series at hexacene. 
Exploring this further, we find that the MP3 contribution to CCSD(T) is not entirely converged for hexacene, and as indicated in Table S1 of the SI, the MP3 basis set size dependence increases with increasing system size. 
By repeating the FPA and extrapolating the MP3 corrections from trends observed in smaller acenes and basis sets, we find a difference of -0.1~eV in the resulting IP of hexacene with respect to Ref.~\citenum{deleuze_benchmark_2003}. 
Our resulting CCSD(T) best theoretical estimates~(BTEs) for the vertical IPs are shown in Table.~\ref{table:ccsdt-ip}. 
Our CCSD(T) calculations are performed with the Gaussian~09~E.01~code\cite{g09} with standard settings, including core electrons in the correlation computation and neglecting relativistic effects as usual. Details of our calculations and analysis of the FPA in Ref.~\citenum{deleuze_benchmark_2003} are provided in the SI. 
In addition, we adopt the EA CCSD(T) reference values of Ref.~\citenum{hajgato_benchmark_2009}.

\begin{table}[h!]
\begin{tabular}{lcc}
\hline\hline
& Ref.~\citenum{deleuze_benchmark_2003} & This work \\
 \\
Benzene      & 9.45 & $9.44 \pm 0.01$\\
Naphthalene  & 8.24 & $8.25 \pm 0.01$\\
Anthracene   & 7.47 & $7.48 \pm 0.03$\\
Tetracene    & 6.95 & $6.96 \pm 0.03$\\
Pentacene    & 6.57 & $6.58 \pm 0.03$\\
Hexacene     & 6.43 & $6.32 \pm 0.03$\\
\hline\hline
\end{tabular}
\caption{Best theoretical estimates based on CCSD(T) calculations following Ref.~\citenum{deleuze_benchmark_2003}, for the vertical IP of the acenes.
Here, we compare our calculations with those of Ref.~\citenum{deleuze_benchmark_2003}, determined by a focal point analysis\cite{deleuze_benchmark_2003} at the CCSD(T) level of theory.
All energies are in units of eV.}
\label{table:ccsdt-ip}
\end{table}

\begin{figure*}[h!]
\begin{tabular}{c|c|c}
\multicolumn{3}{c}{\includegraphics{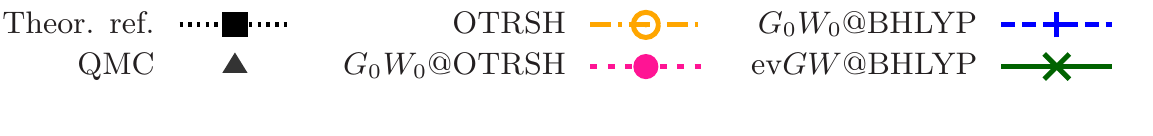}}\\
\hspace{0.4in}IP -- EA gap & \hspace{0.4in}Vertical IP & \hspace{0.4in}Vertical EA\\
\includegraphics[width=0.3\linewidth]{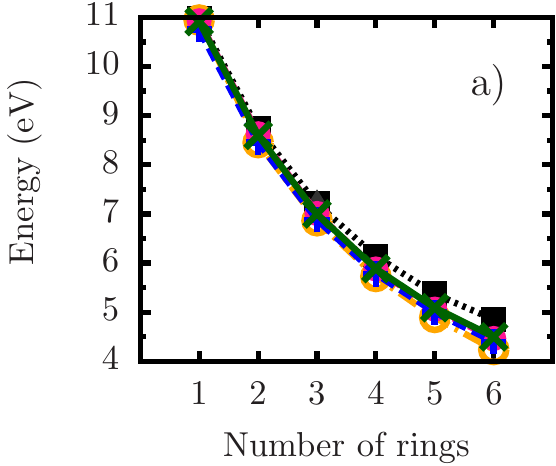}&
\includegraphics[width=0.3\linewidth]{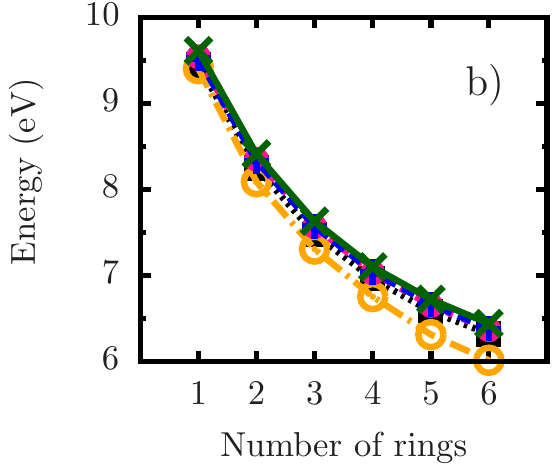}&
\includegraphics[width=0.3\linewidth]{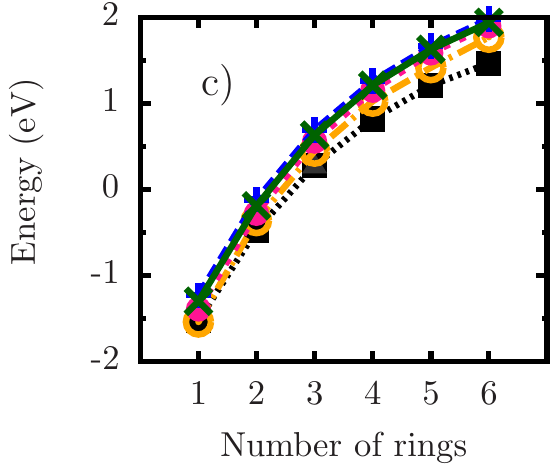}\\
\includegraphics[width=0.3\linewidth]{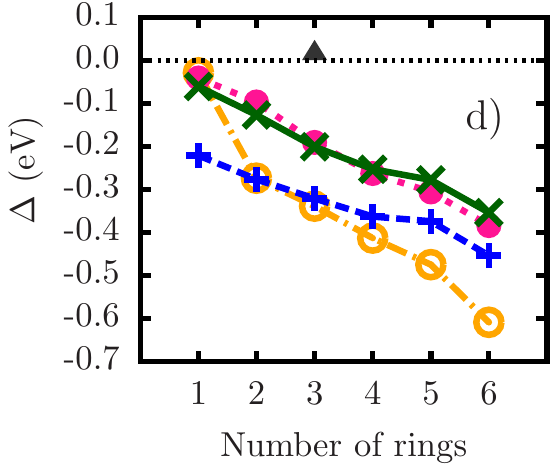}&
\includegraphics[width=0.3\linewidth]{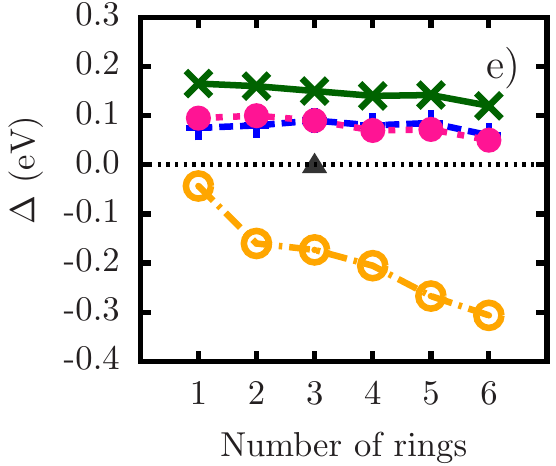}&
\includegraphics[width=0.3\linewidth]{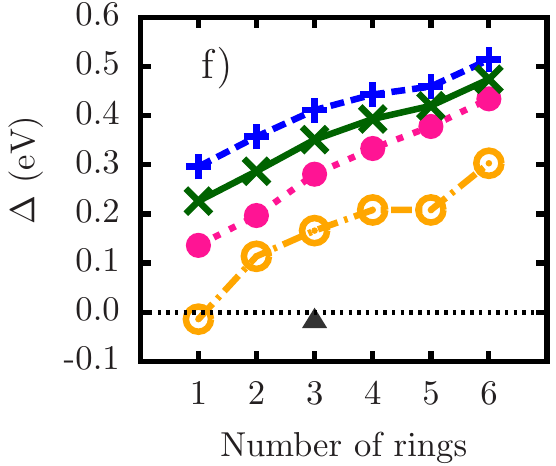}
\end{tabular}
\caption{Charged  excitations of oligoacenes calculated within $GW$ and DFT are compared to CCSD(T); our IPs in Table~\ref{table:ccsdt-ip} and EAs from Refs.~\citenum{hajgato_benchmark_2008}.
Calculated IP--EA gaps, vertical ionization potentials (IP), electron affinities (EA) and their corresponding difference with respect to the theoretical reference, $\Delta$, are shown in panels a - f.
Several $GW$ approaches are  considered (see text).
For comparison, quantum Monte Carlo (QMC) data from Ref.~\citenum{dupuy_vertical_2015} for anthracene are also shown.}
\label{fig:GW-vs-ref}
\end{figure*}
\begin{figure*}
\includegraphics{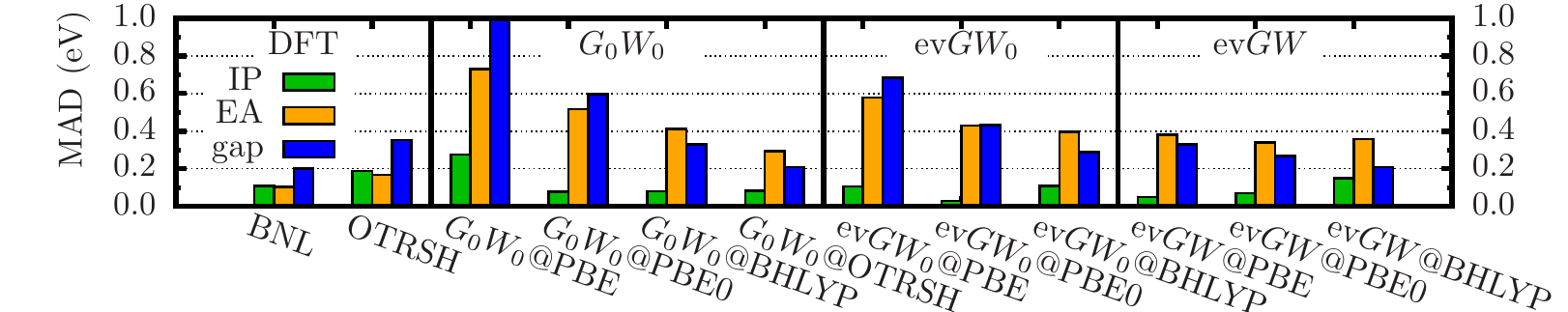}\\
\includegraphics{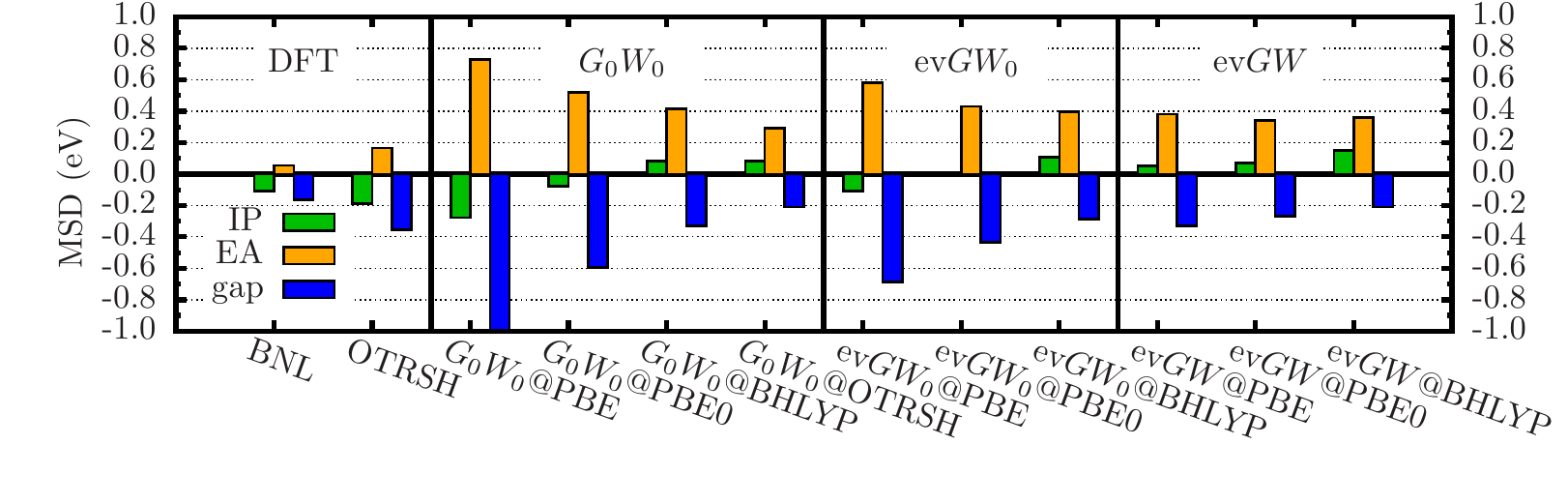}
\caption{
Top: Mean absolute deviation~(MAD) with respect to the theoretical reference [CCSD(T)] in the calculated IPs (green bars), EAs (orange bars) and IP--EA gaps (blue bars) of the acene family of molecules.
Bottom: Mean signed deviation (MSD). Several DFT and $GW$ approximations are considered (see text). 
}
\label{fig:msd}
\end{figure*}

\subsection{Charged excitations with $GW$ and DFT-OTRSH}
\label{sect:gw}

In this section, we present and discuss IPs and EAs calculated using DFT-OTRSH and $GW$, comparing to our CCSD(T) BTEs as defined in the previous section. 
Experimental values for IPs and EAs of the acenes are given in Refs.~\citenum{frank_unoccupied_1988,pope_electronic_1999,nist} and are described as “vertical”; however, recent work\cite{hajgato_benchmark_2008,deleuze_benchmark_2003} with CCSD(T) has suggested that these values are actually “adiabatic” since their best adiabatic estimates match the experimental values within 0.02~eV.  
Note that the naphthalene EA of −0.190~eV measured by electron transmission spectroscopy,\cite{burrow_electron_1987} first ascribed as vertical, is now considered adiabatic due to the presence of vibrational features in the spectrum.\cite{heinis_electron_1993}
For these reasons, we choose not to compare explicitly with experiments in this work and instead benchmark against high-level CCSD(T) calculations: we use our own CCSD(T) IPs, as shown in Section~\ref{sect:ccsdt}, and take CCSD(T) EAs from Ref.~\citenum{hajgato_benchmark_2008}.
We note that vertical IP/EAs of anthracene calculated with QMC are in excellent agreement with our CCSD(T) references.\cite{dupuy_vertical_2015}

In Figure~\ref{fig:GW-vs-ref}, calculated charged-excitations are compared to the CCSD(T) reference data (in black dotted lines and squares).
For clarity, only a few representatives of each $GW$ and DFT scheme are shown;
$G_0W_0$@BHLYP (blue dashed lines and crosses), $G_0W_0$@OTRSH (pink dotted lines and filled circles), ev$GW$@BHLYP (green lines and crosses) and
OTRSH (yellow dashed lines and circles). 
Note that quantum Monte Carlo data\cite{dupuy_vertical_2015} (dark-grey triangles) agree well with the CCSD(T) BTE values for anthracene.
For completeness, the mean signed deviation~($\text{MSD}= 1/N_i \sum_i^{N_i} E_i - E_{\mbox{\scriptsize ref}}$) and mean absolute deviation~($\text{MAD}= 1/N_i \sum_i^{N_i} |E_i - E_{\mbox{\scriptsize ref}}|$) with respect to the CCSD(T) BTEs for all of the approximations considered in this work are shown in Figure~\ref{fig:msd}.

In Figure~\ref{fig:GW-vs-ref}, we plot the IP/EA/gap calculated with OTRSH in yellow circles and dashed-lines; the calculations shown here agree well (within 0.05~eV) with previous works.\cite{refaely-abramson_gap_2013,korzdorfer_long-range_2011,refaely-abramson_quasiparticle_2012}
For benzene, the OTRSH IP and EA are in perfect agreement with the CCSD(T) reference.
However, the agreement deteriorates for larger acenes, in agreement with Ref.~\citenum{korzdorfer_long-range_2011}, possibly due to the fact that as OTRSH is tuned to fulfill the DFT ionization potential theorem, its performance is dependent on the reliability of $\Delta$SCF; larger molecules can show larger frontier-orbital delocalization and $\Delta$SCF is known to perform poorly when orbitals are delocalized, e.g. in the asymptotic limit of infinite molecules and in extended systems due to approximate exchange-correlation potentials.\cite{vlcek_deviations_2015,whittleton_density-functional_2015,mori-sanchez_localization_2008,marsili_ab_2013,gavnholt_delta_2008}
Note that BNL gives the best overall agreement to the reference values, with an MAD of only $\sim$0.1~eV for the IPs, EAs and gaps, as already found in Ref.~\citenum{stein_fundamental_2010}.
OTRSH exhibits a larger MAD, e.g. $\sim$0.3 eV for the IP--EA gap.

We now turn to our results with $G_0W_0$.
First, $G_0W_0$@PBE, severely underestimates the QP gaps, with a MSD of $-1.0$~eV, in agreement with previous findings.\cite{x._blase_first-principles_2011,bruneval_benchmarking_2013,marom_benchmark_2012,van_setten_gw100:_2015,korzdorfer_strategy_2012,korzdorfer_strategy_2012}
The shortcomings of $G_0W_0$@PBE are well-known and discussed in Refs.~\citenum{bruneval_benchmarking_2013,x._blase_first-principles_2011,marom_benchmark_2012,korzdorfer_strategy_2012}.
Note that standard $G_0W_0$ calculations of charged excitations of the acenes have been reported using plane-wave approaches,\cite{sharifzadeh_quantitative_2012,refaely-abramson_gap_2013,lischner_effects_2014,van_setten_gw100:_2015,rangel_structural_2016} and the level of convergence and the nature of the frequency-integration schemes can lead to qualitative differences from the work presented here that are well documented.\cite{sharifzadeh_quantitative_2012,van_setten_gw100:_2015} 

One known strategy to improve over $G_0W_0$@PBE is to use an XC functional with a fraction of exact exchange as starting point.\cite{bruneval_benchmarking_2013,marom_benchmark_2012}
We find that the global hybrid providing the best results (lower MAD) is $G_0W_0$@BHLYP, which has an MAD of $0.3$~eV in the calculated IP--EA acene gaps (See Figure~\ref{fig:msd}), good accuracy at a reasonable computational cost.
Notably, the OTRSH starting point leads to highly accurate QP energies, with an MAD for the IP--EA gap of only $0.2$~eV, and in close agreement with more expensive ev$GW$ schemes (see Figure~\ref{fig:GW-vs-ref}).

The excellent performance of $G_0W_0$@OTRSH, as hypothesized in Ref.~\citenum{refaely-abramson_quasiparticle_2012}, is consistent with the conclusions of Gallandi and K\"orzd\"orfer\cite{gallandi_long-range_2015} who explored several $GW$ approaches and found that a tuned long-range separated hybrid (namely the IP-tuned LC-$\omega$PBE\cite{korzdorfer_long-range_2011}, equivalent to OTRSH with $\alpha=0$ and $\beta=1$) $G_0W_0$ starting point yields charged excitation energies within 0.1~eV of experiment and ev$GW$ in a set of molecules that includes some (but not all) of the acenes.
Their work was extended in Refs.~\citenum{knight_accurate_2016} and \citenum{gallandi_accurate_2016} where a CCSD(T) reference was used, and where it was reported that the LC-$\omega$PBE starting point leads to the smallest MAD~(0.2~eV for EAs and 0.1 for IPs) in a set of short- to medium- sized molecules.
In agreement with the findings of Ref.~\citenum{gallandi_accurate_2016}, the RSH starting point for $G_0W_0$ with fixed $\alpha$ and $\gamma$ parameters can lead to the same level of accuracy than $G_0W_0@$OTRSH in the IP and EA energy levels of the acene molecules (with a MAD of $\leq$ 0.3), see Sect. 2 of the SI for details.
Prior calculations,\cite{korzdorfer_strategy_2012,marom_benchmark_2012}
including some acene molecules, report that the PBE0 starting point provides the best overall  QP energies relative to photoemission experiments along a broad energy range; here, we compare to CCSD(T) and focus only on the frontier molecular levels.

A second approach known to provide accurate QP energies is eigenvalue self-consistency.\cite{x._blase_first-principles_2011} 
Here, we test two different levels of eigenvalue self-consistency: partial self-consistency, updating eigenvalues only in $G$ (ev$GW_0$); and full self-consistency, updating eigenvalues in both $G$ and $W$ (ev$GW$).
We find that ev$GW_0$ leads to unsatisfactory results for these molecules, 
unless $W_0$ from BHLYP is used,
e.g., ev$GW_0$@PBE, ev$GW_0$PBE0 and ev$GW_0$BHLYP result in a MAD of $0.7$, $0.4$  and $0.3$~eV, respectively, for the IP--EA gap.
Moreover, no clear improvement is found with ev$GW_0$ over $G_0W_0$@BHLYP; the two approaches result in nearly equivalent QP energies (within $0.05$~eV).
On the other hand, ev$GW$ results in overall good agreement to the reference values, with an MAD of $\sim 0.2$~eV for the QP gap (see Figure~\ref{fig:msd}).
We also highlight that there is not much spread in the ev$GW$ QP energies with respect to the DFT starting point; in fact, the ev$GW$ gap of benzene is predicted to be $10.9$~eV independent of the DFT starting points considered here.
For larger molecules, ev$GW$ with different starting points can, in some cases, lead to more appreciable differences: for example, a difference of $0.2$~eV in QP is observed with PBE or BHLYP starting points for tetracene.
In Ref.~\citenum{marom_benchmark_2012} by  considering the extreme starting-points, PBE and HF (with 0\% and 100\% exact exchange, respectively), a larger difference ($\sim 0.4$ eV) was found in the resulting ev$GW$ gaps of organic molecules. 
Nevertheless, the starting-point dependence of ev$GW$ is less than in the case of $G_0W_0$, which is typically $\sim 1.4$~eV for aromatic molecules.\cite{marom_benchmark_2012} Hence, the ev$GW$ method is an attractive approach due to its relatively-minimal starting-point dependence and good accuracy, in spite of its higher cost with respect to one-shot $G_0W_0$.

\begin{figure}
\includegraphics{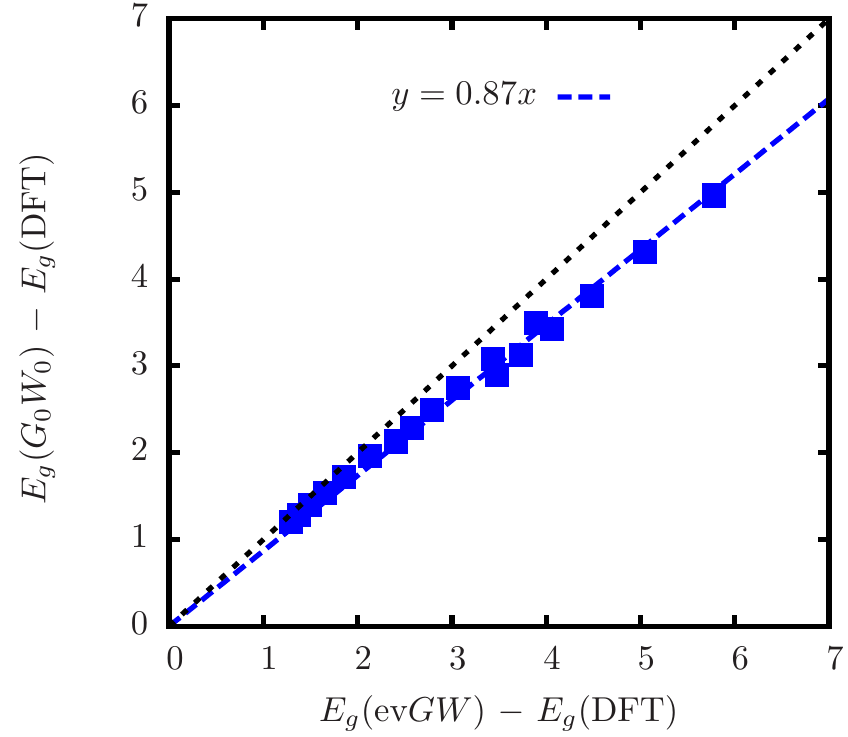}
\caption{\small Correlation and linear fit between ev$GW$ gaps and their corresponding $G_0W_0$ gaps for the oligoacenes.
Results of $GW$ calculations with PBE, PBE0 and BHLYP starting points are used in constructing this plot.
All energies are in units of eV.
}
\label{fig:evgw-g0w0}
\end{figure}

The ev$GW$ and $G_0W_0$ approaches and their corresponding self-energy corrections are linearly correlated. 
In Figure~\ref{fig:evgw-g0w0}, we show corrections to the IP--EA gap [gap($GW$) $-$ gap(DFT)] obtained from both $G_0W_0$ and ev$GW$.
As expected, ev$GW$ leads to larger gaps than $G_0W_0$@PBE.\cite{x._blase_first-principles_2011,korzdorfer_long-range_2011,knight_accurate_2016}
Interestingly, independent of starting point, we find that our $G_0W_0$ corrections are consistently 87\% of the corresponding ev$GW$ corrections (see dashed blue line in Fig.~\ref{fig:evgw-g0w0} with a slope of $0.87$ and a standard deviation of $< 0.01$~eV).
Note that six points between 4 and 7 eV lie slightly below this linear trend (dashed blue line); these points use a PBE starting point and are best fit with a slightly smaller slope of $0.85$ (not shown). 
This simple relation, consistent with the tendency of $G_0W_0$ to underestimate gaps due to over screening\cite{x._blase_first-principles_2011} and the fact that the screening is similar enough across the acene series, would allow for an accurate estimation of the ev$GW$ gap from $G_0W_0$ corrections for acenes, or even of the $G_0W_0$ gap from other DFT starting points.

In agreement with recent work, \cite{korzdorfer_long-range_2011,refaely-abramson_quasiparticle_2012,knight_accurate_2016,gallandi_accurate_2016} we find that RSH or $GW$ can provide highly accurate frontier orbital energies for small acenes; for benzene, OTRSH and $GW$@OTRSH gives IPs and EAs within 0.1~eV of the CCSD(T) reference. 
For medium-sized molecules the accuracy of both RSH and $GW$ decreases; for hexacene OTRSH presents deviations of $\sim$0.3~eV in both the IP and the EA, adding to an MAD of $\sim$0.6~eV in the gap; the more accurate $GW$ approaches tested in this work ($GW$@OTRSH, $GW$@BHLYP and ev$GW$) predict IPs within 0.1~eV but can overestimate EAs by up to $0.4-0.5$~eV (see Fig.~\ref{fig:GW-vs-ref}).
Nevertheless, since the deviation grows linearly with the number of rings~($N_{\mbox{\scriptsize ring}}$), the EAs can be linearly extrapolated from the $GW$ results as:
\begin{equation}
\mbox{EA} = \mbox{EA}^{\mbox{\scriptsize $GW$@OTRSH}} - (0.087\,\, \mbox{eV}) N_{\mbox{\scriptsize ring}} + 0.06\,\, \mbox{eV},
\end{equation}
where we have subtracted a linear function of the number of rings from the $GW$@OTRSH EA energies, and obtained the EA energies within 0.02~eV of the CCSD(T) reference.
This simple relation, though effective here, may not be transferable to other aromatic hydrocarbon families, and in general will not be applicable for any other non-ringed system; $GW$ calculations beyond the approaches used in this work may yield further insight into this trend.
It is worth noticing that wavefunctions of larger acenes have some multi-reference character,\cite{bendikov_oligoacenes:_2004} which might explain the observed limited performance of $GW$ in the large molecule regime.

In this work we fix the fraction of short-range exact exchange in all DFT-RSH calculations; thus one straightforward extension is to tune the $\alpha$ parameter  with theorems of DFT, along the lines of Refs.\citenum{srebro_does_2012}, \citenum{refaely-abramson_quasiparticle_2012} and \citenum{luftner_experimental_2014}.
Tuning the short-range HF parameter may lead to a better description of systems with localized electrons; nevertheless, for the acenes, with only $s$ and $p$ electrons, it is not evident a priori that such tuning would improve the accuracy of both OTRSH and $GW$ based on an OTRSH starting point.
In fact, for benzene and pentacene, setting $\alpha$ to either 0.0 or 0.2 leads to negligible changes (by $<0.1$~eV) in the IPs or EAs, as shown in Refs.~\citenum{refaely-abramson_quasiparticle_2012} and \citenum{egger_outer-valence_2014}.
Further, when it comes to the $\pi$ and $\pi^*$ orbital energies of representative organic molecules, it has been shown in Ref.~\citenum{gallandi_accurate_2016} that the tuning of the $\alpha$ parameter does not significantly affect the performance of OTRSH or $G_0W_0$@OTRSH.
Beyond eigenvalue self-consistent~$GW$, total energy differences from $GW$\cite{bruneval_$gw$_2009,caruso_bond_2013} might also result in more accurate frontier orbital energies. 
We leave these considerations to be explored in future work.

In summary, our results indicate that going beyond standard $G_0W_0$ is crucial to achieve CCSD(T) accuracy; including self-consistency, such as in the ev$GW$ method, or adding a fraction of exact exchange are both successful strategies for describing charged-excitations. 
Notably, using OTRSH as a starting point for $G_0W_0$ provides highly accurate energetics relative to CCSD(T), in agreement with the more expensive ev$GW$ results.

\section{Conclusions}
\label{sect:conclusions}

In this work, we have calculated IPs and EAs of acene molecules with DFT, $GW$, and wavefunction-based approaches. 
We have built upon and extended the CCSD(T) reference data of Ref.~\citenum{deleuze_benchmark_2003} for IPs of the larger acenes. 
Using this new CCSD(T) reference, we have benchmarked $GW$ under several approximations and DFT with range-separated hybrid methods and found that both $G_0W_0$@OTRSH and ev$GW$ consistently perform well, yielding quantitative IP energies within 0.1 eV of CCSD(T) across the acene series.
Nevertheless, all $GW$ approaches studied here lead to qualitative deviations for the larger acenes, 
suggesting the need to go beyond eigenvalue-self-consistent $GW$ methods to do better.
Moreover, we have found that DFT with OTRSH or BNL functionals can perform as well as the most effective $GW$ approaches for benzene, but their resulting IPs and EAs deteriorate as the molecules get larger in the series, a behavior attributable to the known deficiencies of $\Delta$SCF in the asymptotic limit of large molecules towards extended systems.

\begin{acknowledgement}
T. Rangel thanks S. Refaely-Abramson for her valuable feedback and discussions on the OTRSH functional. 
F. Bruneval acknowledges the Enhanced Eurotalent program and the France Berkeley Fund for supporting his sabbatical leave in UC Berkeley. 
This research was supported by the SciDAC Program on Excited State Phenomena in Energy Materials funded by the U.S. Department of Energy, Office of Basic Energy Sciences and of Advanced Scientific Computing Research, under Contract No. DE-AC02-05CH11231 at Lawrence Berkeley National Laboratory. 
Work at the Molecular Foundry was supported by the Office of Science, Office of Basic Energy Sciences, of the U.S. Department of Energy under Contract No. DE-AC02-05CH11231. 
This research used resources of the National Energy Research Scientific Computing Center, which is supported by the Office of Science of the U.S. Department of Energy.
\end{acknowledgement}

\begin{suppinfo}
We tabulate all charged excitation energies calculated in this work, we provide the relaxed geometries used in this work, and we give a detailed explanation on the way we obtain the revised CCSD(T) data.
\end{suppinfo}

\bibliography{paper}

\providecommand{\latin}[1]{#1}
\providecommand*\mcitethebibliography{\thebibliography}
\csname @ifundefined\endcsname{endmcitethebibliography}
  {\let\endmcitethebibliography\endthebibliography}{}
\begin{mcitethebibliography}{91}
\providecommand*\natexlab[1]{#1}
\providecommand*\mciteSetBstSublistMode[1]{}
\providecommand*\mciteSetBstMaxWidthForm[2]{}
\providecommand*\mciteBstWouldAddEndPuncttrue
  {\def\EndOfBibitem{\unskip.}}
\providecommand*\mciteBstWouldAddEndPunctfalse
  {\let\EndOfBibitem\relax}
\providecommand*\mciteSetBstMidEndSepPunct[3]{}
\providecommand*\mciteSetBstSublistLabelBeginEnd[3]{}
\providecommand*\EndOfBibitem{}
\mciteSetBstSublistMode{f}
\mciteSetBstMaxWidthForm{subitem}{(\alph{mcitesubitemcount})}
\mciteSetBstSublistLabelBeginEnd
  {\mcitemaxwidthsubitemform\space}
  {\relax}
  {\relax}

\bibitem[Anthony(2006)]{anthony_functionalized_2006}
Anthony,~J.~E. \emph{Chem. Rev.} \textbf{2006}, \emph{106}, 5028\relax
\mciteBstWouldAddEndPuncttrue
\mciteSetBstMidEndSepPunct{\mcitedefaultmidpunct}
{\mcitedefaultendpunct}{\mcitedefaultseppunct}\relax
\EndOfBibitem
\bibitem[Forrest and Thompson(2007)Forrest, and
  Thompson]{forrest_introduction:_2007}
Forrest,~S.~R.; Thompson,~M.~E. \emph{Chem. Rev.} \textbf{2007}, \emph{107},
  923--925\relax
\mciteBstWouldAddEndPuncttrue
\mciteSetBstMidEndSepPunct{\mcitedefaultmidpunct}
{\mcitedefaultendpunct}{\mcitedefaultseppunct}\relax
\EndOfBibitem
\bibitem[Br\'edas \latin{et~al.}(2009)Br\'edas, Norton, Cornil, and
  Coropceanu]{bredas_molecular_2009}
Br\'edas,~J.-L.; Norton,~J.~E.; Cornil,~J.; Coropceanu,~V. \emph{Acc. Chem.
  Res.} \textbf{2009}, \emph{42}, 1691--1699\relax
\mciteBstWouldAddEndPuncttrue
\mciteSetBstMidEndSepPunct{\mcitedefaultmidpunct}
{\mcitedefaultendpunct}{\mcitedefaultseppunct}\relax
\EndOfBibitem
\bibitem[Smith and Michl(2010)Smith, and Michl]{smith_singlet_2010}
Smith,~M.~B.; Michl,~J. \emph{Chem. Rev.} \textbf{2010}, \emph{110}, 6891\relax
\mciteBstWouldAddEndPuncttrue
\mciteSetBstMidEndSepPunct{\mcitedefaultmidpunct}
{\mcitedefaultendpunct}{\mcitedefaultseppunct}\relax
\EndOfBibitem
\bibitem[Lee \latin{et~al.}(2013)Lee, Jadhav, Reusswig, Yost, Thompson,
  Congreve, Hontz, Van~Voorhis, and Baldo]{lee_singlet_2013}
Lee,~J.; Jadhav,~P.; Reusswig,~P.~D.; Yost,~S.~R.; Thompson,~N.~J.;
  Congreve,~D.~N.; Hontz,~E.; Van~Voorhis,~T.; Baldo,~M.~A. \emph{Acc. Chem.
  Res.} \textbf{2013}, \emph{46}, 1300--1311\relax
\mciteBstWouldAddEndPuncttrue
\mciteSetBstMidEndSepPunct{\mcitedefaultmidpunct}
{\mcitedefaultendpunct}{\mcitedefaultseppunct}\relax
\EndOfBibitem
\bibitem[Battersby(2004)]{_space_}
Battersby,~S. Space molecules point to organic origins. 2004;
  \url{https://www.newscientist.com/article/dn4552-space-molecules-point-to-organic-origins/},
  (Visited on 03/09/2016).\relax
\mciteBstWouldAddEndPunctfalse
\mciteSetBstMidEndSepPunct{\mcitedefaultmidpunct}
{}{\mcitedefaultseppunct}\relax
\EndOfBibitem
\bibitem[Mulas \latin{et~al.}(2006)Mulas, Malloci, Joblin, and
  Toublanc]{mulas_estimated_2006}
Mulas,~G.; Malloci,~G.; Joblin,~C.; Toublanc,~D. \emph{Astron. Astrophys.}
  \textbf{2006}, \emph{446}, 13\relax
\mciteBstWouldAddEndPuncttrue
\mciteSetBstMidEndSepPunct{\mcitedefaultmidpunct}
{\mcitedefaultendpunct}{\mcitedefaultseppunct}\relax
\EndOfBibitem
\bibitem[Boersma \latin{et~al.}(2014)Boersma, C.~W.~Bauschlicher, Ricca,
  Mattioda, Cami, Peeters, Armas, Saborido, Hudgins, and
  Allamandola]{boersma_nasa_2014}
Boersma,~C.; C.~W.~Bauschlicher,~J.; Ricca,~A.; Mattioda,~A.~L.; Cami,~J.;
  Peeters,~E.; Armas,~F. S.~d.; Saborido,~G.~P.; Hudgins,~D.~M.;
  Allamandola,~L.~J. \emph{Astrophys. J. Suppl. Ser.} \textbf{2014},
  \emph{211}, 8\relax
\mciteBstWouldAddEndPuncttrue
\mciteSetBstMidEndSepPunct{\mcitedefaultmidpunct}
{\mcitedefaultendpunct}{\mcitedefaultseppunct}\relax
\EndOfBibitem
\bibitem[Hajgat\'o \latin{et~al.}(2008)Hajgat\'o, Deleuze, Tozer, and
  De~Proft]{hajgato_benchmark_2008}
Hajgat\'o,~B.; Deleuze,~M.~S.; Tozer,~D.~J.; De~Proft,~F. \emph{J. Chem. Phys}
  \textbf{2008}, \emph{129}, 084308\relax
\mciteBstWouldAddEndPuncttrue
\mciteSetBstMidEndSepPunct{\mcitedefaultmidpunct}
{\mcitedefaultendpunct}{\mcitedefaultseppunct}\relax
\EndOfBibitem
\bibitem[Deleuze \latin{et~al.}(2003)Deleuze, Claes, Kryachko, and
  Fran\c{c}ois]{deleuze_benchmark_2003}
Deleuze,~M.~S.; Claes,~L.; Kryachko,~E.~S.; Fran\c{c}ois,~J.-P. \emph{J. Chem.
  Phys.} \textbf{2003}, \emph{119}, 3106\relax
\mciteBstWouldAddEndPuncttrue
\mciteSetBstMidEndSepPunct{\mcitedefaultmidpunct}
{\mcitedefaultendpunct}{\mcitedefaultseppunct}\relax
\EndOfBibitem
\bibitem[Dupuy \latin{et~al.}(2015)Dupuy, Bouaouli, Mauri, Sorella, and
  Casula]{dupuy_vertical_2015}
Dupuy,~N.; Bouaouli,~S.; Mauri,~F.; Sorella,~S.; Casula,~M. \emph{J. Chem.
  Phys.} \textbf{2015}, \emph{142}, 214109\relax
\mciteBstWouldAddEndPuncttrue
\mciteSetBstMidEndSepPunct{\mcitedefaultmidpunct}
{\mcitedefaultendpunct}{\mcitedefaultseppunct}\relax
\EndOfBibitem
\bibitem[Burrow \latin{et~al.}(1987)Burrow, Michejda, and
  Jordan]{burrow_electron_1987}
Burrow,~P.~D.; Michejda,~J.~A.; Jordan,~K.~D. \emph{J. Chem. Phys.}
  \textbf{1987}, \emph{86}, 9\relax
\mciteBstWouldAddEndPuncttrue
\mciteSetBstMidEndSepPunct{\mcitedefaultmidpunct}
{\mcitedefaultendpunct}{\mcitedefaultseppunct}\relax
\EndOfBibitem
\bibitem[Tamblyn \latin{et~al.}(2014)Tamblyn, Refaely-Abramson, Neaton, and
  Kronik]{tamblyn_simultaneous_2014}
Tamblyn,~I.; Refaely-Abramson,~S.; Neaton,~J. f.~B.; Kronik,~L. \emph{J. Phys.
  Chem. Lett.} \textbf{2014}, \emph{5}, 2734--2741\relax
\mciteBstWouldAddEndPuncttrue
\mciteSetBstMidEndSepPunct{\mcitedefaultmidpunct}
{\mcitedefaultendpunct}{\mcitedefaultseppunct}\relax
\EndOfBibitem
\bibitem[Hedin(1965)]{hedin_new_1965}
Hedin,~L. \emph{Phys. Rev.} \textbf{1965}, \emph{139}, A796--A823\relax
\mciteBstWouldAddEndPuncttrue
\mciteSetBstMidEndSepPunct{\mcitedefaultmidpunct}
{\mcitedefaultendpunct}{\mcitedefaultseppunct}\relax
\EndOfBibitem
\bibitem[Bruneval(2009)]{bruneval_$gw$_2009}
Bruneval,~F. \emph{Phys. Rev. Lett.} \textbf{2009}, \emph{103}, 176403\relax
\mciteBstWouldAddEndPuncttrue
\mciteSetBstMidEndSepPunct{\mcitedefaultmidpunct}
{\mcitedefaultendpunct}{\mcitedefaultseppunct}\relax
\EndOfBibitem
\bibitem[{X. Blase} \latin{et~al.}(2011){X. Blase}, Attaccalite, and
  Olevano]{x._blase_first-principles_2011}
{X. Blase},; Attaccalite,~C.; Olevano,~V. \emph{Phys. Rev. B} \textbf{2011},
  \emph{83}, 115103\relax
\mciteBstWouldAddEndPuncttrue
\mciteSetBstMidEndSepPunct{\mcitedefaultmidpunct}
{\mcitedefaultendpunct}{\mcitedefaultseppunct}\relax
\EndOfBibitem
\bibitem[Marom \latin{et~al.}(2012)Marom, Caruso, Ren, Hofmann, K\"orzd\"orfer,
  Chelikowsky, Rubio, Scheffler, and Rinke]{marom_benchmark_2012}
Marom,~N.; Caruso,~F.; Ren,~X.; Hofmann,~O.~T.; K\"orzd\"orfer,~T.;
  Chelikowsky,~J.~R.; Rubio,~A.; Scheffler,~M.; Rinke,~P. \emph{Phys. Rev. B}
  \textbf{2012}, \emph{86}, 245127\relax
\mciteBstWouldAddEndPuncttrue
\mciteSetBstMidEndSepPunct{\mcitedefaultmidpunct}
{\mcitedefaultendpunct}{\mcitedefaultseppunct}\relax
\EndOfBibitem
\bibitem[K\"orzd\"orfer and Marom(2012)K\"orzd\"orfer, and
  Marom]{korzdorfer_strategy_2012}
K\"orzd\"orfer,~T.; Marom,~N. \emph{Phys. Rev. B} \textbf{2012}, \emph{86},
  041110\relax
\mciteBstWouldAddEndPuncttrue
\mciteSetBstMidEndSepPunct{\mcitedefaultmidpunct}
{\mcitedefaultendpunct}{\mcitedefaultseppunct}\relax
\EndOfBibitem
\bibitem[Gallandi and K\"orzd\"orfer(2015)Gallandi, and
  K\"orzd\"orfer]{gallandi_long-range_2015}
Gallandi,~L.; K\"orzd\"orfer,~T. \emph{J. Chem. Theory Comput.} \textbf{2015},
  \emph{11}, 5391\relax
\mciteBstWouldAddEndPuncttrue
\mciteSetBstMidEndSepPunct{\mcitedefaultmidpunct}
{\mcitedefaultendpunct}{\mcitedefaultseppunct}\relax
\EndOfBibitem
\bibitem[Knight \latin{et~al.}(2016)Knight, Wang, Gallandi, Dolgounitcheva,
  Ren, Ortiz, Rinke, K\"orzd\"orfer, and Marom]{knight_accurate_2016}
Knight,~J.~W.; Wang,~X.; Gallandi,~L.; Dolgounitcheva,~O.; Ren,~X.;
  Ortiz,~J.~V.; Rinke,~P.; K\"orzd\"orfer,~T.; Marom,~N. \emph{J. Chem. Theory
  Comput.} \textbf{2016}, \emph{12}, 615--626\relax
\mciteBstWouldAddEndPuncttrue
\mciteSetBstMidEndSepPunct{\mcitedefaultmidpunct}
{\mcitedefaultendpunct}{\mcitedefaultseppunct}\relax
\EndOfBibitem
\bibitem[Rangel \latin{et~al.}(2016)Rangel, Berland, Sharifzadeh,
  Brown-Altvater, Lee, Hyldgaard, Kronik, and Neaton]{rangel_structural_2016}
Rangel,~T.; Berland,~K.; Sharifzadeh,~S.; Brown-Altvater,~F.; Lee,~K.;
  Hyldgaard,~P.; Kronik,~L.; Neaton,~J.~B. \emph{Phys. Rev. B} \textbf{2016},
  \emph{93}, 115206\relax
\mciteBstWouldAddEndPuncttrue
\mciteSetBstMidEndSepPunct{\mcitedefaultmidpunct}
{\mcitedefaultendpunct}{\mcitedefaultseppunct}\relax
\EndOfBibitem
\bibitem[Kronik \latin{et~al.}(2012)Kronik, Stein, Refaely-Abramson, and
  Baer]{kronik_excitation_2012}
Kronik,~L.; Stein,~T.; Refaely-Abramson,~S.; Baer,~R. \emph{J. Chem. Theory
  Comput.} \textbf{2012}, \emph{8}, 1515, see references therein.\relax
\mciteBstWouldAddEndPunctfalse
\mciteSetBstMidEndSepPunct{\mcitedefaultmidpunct}
{}{\mcitedefaultseppunct}\relax
\EndOfBibitem
\bibitem[Ziegler(1991)]{ziegler_approximate_1991}
Ziegler,~T. \emph{Chem. Rev.} \textbf{1991}, \emph{91}, 651--667\relax
\mciteBstWouldAddEndPuncttrue
\mciteSetBstMidEndSepPunct{\mcitedefaultmidpunct}
{\mcitedefaultendpunct}{\mcitedefaultseppunct}\relax
\EndOfBibitem
\bibitem[Cohen \latin{et~al.}(2008)Cohen, Mori-S\'anchez, and
  Yang]{cohen_insights_2008}
Cohen,~A.~J.; Mori-S\'anchez,~P.; Yang,~W. \emph{Science} \textbf{2008},
  \emph{321}, 792--794\relax
\mciteBstWouldAddEndPuncttrue
\mciteSetBstMidEndSepPunct{\mcitedefaultmidpunct}
{\mcitedefaultendpunct}{\mcitedefaultseppunct}\relax
\EndOfBibitem
\bibitem[Vl\v{c}ek \latin{et~al.}(2015)Vl\v{c}ek, Eisenberg, Steinle-Neumann,
  Kronik, and Baer]{vlcek_deviations_2015}
Vl\v{c}ek,~V.; Eisenberg,~H.~R.; Steinle-Neumann,~G.; Kronik,~L.; Baer,~R.
  \emph{J. Chem. Phys.} \textbf{2015}, \emph{142}, 034107\relax
\mciteBstWouldAddEndPuncttrue
\mciteSetBstMidEndSepPunct{\mcitedefaultmidpunct}
{\mcitedefaultendpunct}{\mcitedefaultseppunct}\relax
\EndOfBibitem
\bibitem[Gallandi \latin{et~al.}(2016)Gallandi, Marom, Rinke, and
  K\"orzd\"orfer]{gallandi_accurate_2016}
Gallandi,~L.; Marom,~N.; Rinke,~P.; K\"orzd\"orfer,~T. \emph{J. Chem. Theory
  Comput.} \textbf{2016}, \emph{12}, 605\relax
\mciteBstWouldAddEndPuncttrue
\mciteSetBstMidEndSepPunct{\mcitedefaultmidpunct}
{\mcitedefaultendpunct}{\mcitedefaultseppunct}\relax
\EndOfBibitem
\bibitem[Refaely-Abramson \latin{et~al.}(2011)Refaely-Abramson, Baer, and
  Kronik]{refaely-abramson_fundamental_2011}
Refaely-Abramson,~S.; Baer,~R.; Kronik,~L. \emph{Phys. Rev. B} \textbf{2011},
  \emph{84}, 075144\relax
\mciteBstWouldAddEndPuncttrue
\mciteSetBstMidEndSepPunct{\mcitedefaultmidpunct}
{\mcitedefaultendpunct}{\mcitedefaultseppunct}\relax
\EndOfBibitem
\bibitem[K\"orzd\"orfer \latin{et~al.}(2011)K\"orzd\"orfer, Sears, Sutton, and
  Br\'edas]{korzdorfer_long-range_2011}
K\"orzd\"orfer,~T.; Sears,~J.~S.; Sutton,~C.; Br\'edas,~J.-L. \emph{J. Chem.
  Phys.} \textbf{2011}, \emph{135}, 204107\relax
\mciteBstWouldAddEndPuncttrue
\mciteSetBstMidEndSepPunct{\mcitedefaultmidpunct}
{\mcitedefaultendpunct}{\mcitedefaultseppunct}\relax
\EndOfBibitem
\bibitem[Refaely-Abramson \latin{et~al.}(2012)Refaely-Abramson, Sharifzadeh,
  Govind, Autschbach, Neaton, Baer, and
  Kronik]{refaely-abramson_quasiparticle_2012}
Refaely-Abramson,~S.; Sharifzadeh,~S.; Govind,~N.; Autschbach,~J.;
  Neaton,~J.~B.; Baer,~R.; Kronik,~L. \emph{Phys. Rev. Lett.} \textbf{2012},
  \emph{109}, 226405\relax
\mciteBstWouldAddEndPuncttrue
\mciteSetBstMidEndSepPunct{\mcitedefaultmidpunct}
{\mcitedefaultendpunct}{\mcitedefaultseppunct}\relax
\EndOfBibitem
\bibitem[Refaely-Abramson \latin{et~al.}(2013)Refaely-Abramson, Sharifzadeh,
  Jain, Baer, Neaton, and Kronik]{refaely-abramson_gap_2013}
Refaely-Abramson,~S.; Sharifzadeh,~S.; Jain,~M.~a.; Baer,~R.; Neaton,~J.~B.;
  Kronik,~L. \emph{Phys. Rev. B} \textbf{2013}, \emph{88}, 081204\relax
\mciteBstWouldAddEndPuncttrue
\mciteSetBstMidEndSepPunct{\mcitedefaultmidpunct}
{\mcitedefaultendpunct}{\mcitedefaultseppunct}\relax
\EndOfBibitem
\bibitem[Egger \latin{et~al.}(2014)Egger, Weissman, Refaely-Abramson,
  Sharifzadeh, Dauth, Baer, K\"ummel, Neaton, Zojer, and
  Kronik]{egger_outer-valence_2014}
Egger,~D.~A.; Weissman,~S.; Refaely-Abramson,~S.; Sharifzadeh,~S.; Dauth,~M.;
  Baer,~R.; K\"ummel,~S.; Neaton,~J.~B.; Zojer,~E.; Kronik,~L. \emph{J. Chem.
  Theory Comput.} \textbf{2014}, \emph{10}, 1934--1952\relax
\mciteBstWouldAddEndPuncttrue
\mciteSetBstMidEndSepPunct{\mcitedefaultmidpunct}
{\mcitedefaultendpunct}{\mcitedefaultseppunct}\relax
\EndOfBibitem
\bibitem[Baer and Neuhauser(2005)Baer, and Neuhauser]{baer_density_2005}
Baer,~R.; Neuhauser,~D. \emph{Phys. Rev. Lett.} \textbf{2005}, \emph{94},
  043002\relax
\mciteBstWouldAddEndPuncttrue
\mciteSetBstMidEndSepPunct{\mcitedefaultmidpunct}
{\mcitedefaultendpunct}{\mcitedefaultseppunct}\relax
\EndOfBibitem
\bibitem[Livshits and Baer(2007)Livshits, and
  Baer]{livshits_well-tempered_2007}
Livshits,~E.; Baer,~R. \emph{Phys. Chem. Chem. Phys.} \textbf{2007}, \emph{9},
  2932--2941\relax
\mciteBstWouldAddEndPuncttrue
\mciteSetBstMidEndSepPunct{\mcitedefaultmidpunct}
{\mcitedefaultendpunct}{\mcitedefaultseppunct}\relax
\EndOfBibitem
\bibitem[Kuritz \latin{et~al.}(2011)Kuritz, Stein, Baer, and
  Kronik]{kuritz_charge-transfer-like_2011}
Kuritz,~N.; Stein,~T.; Baer,~R.; Kronik,~L. \emph{J. Chem. Theory Comput.}
  \textbf{2011}, \emph{7}, 2408--2415\relax
\mciteBstWouldAddEndPuncttrue
\mciteSetBstMidEndSepPunct{\mcitedefaultmidpunct}
{\mcitedefaultendpunct}{\mcitedefaultseppunct}\relax
\EndOfBibitem
\bibitem[Leininger \latin{et~al.}(1997)Leininger, Stoll, Werner, and
  Savin]{leininger_combining_1997}
Leininger,~T.; Stoll,~H.; Werner,~H.-J.; Savin,~A. \emph{Chem. Phys. Lett.}
  \textbf{1997}, \emph{275}, 151--160\relax
\mciteBstWouldAddEndPuncttrue
\mciteSetBstMidEndSepPunct{\mcitedefaultmidpunct}
{\mcitedefaultendpunct}{\mcitedefaultseppunct}\relax
\EndOfBibitem
\bibitem[Yanai \latin{et~al.}(2004)Yanai, Tew, and Handy]{yanai_new_2004}
Yanai,~T.; Tew,~D.~P.; Handy,~N.~C. \emph{Chem. Phys. Lett.} \textbf{2004},
  \emph{393}, 51--57\relax
\mciteBstWouldAddEndPuncttrue
\mciteSetBstMidEndSepPunct{\mcitedefaultmidpunct}
{\mcitedefaultendpunct}{\mcitedefaultseppunct}\relax
\EndOfBibitem
\bibitem[Perdew \latin{et~al.}(1996)Perdew, Burke, and
  Ernzerhof]{perdew_generalized_1996}
Perdew,~J.~P.; Burke,~K.; Ernzerhof,~M. \emph{Phys. Rev. Lett.} \textbf{1996},
  \emph{77}, 3865\relax
\mciteBstWouldAddEndPuncttrue
\mciteSetBstMidEndSepPunct{\mcitedefaultmidpunct}
{\mcitedefaultendpunct}{\mcitedefaultseppunct}\relax
\EndOfBibitem
\bibitem[Srebro and Autschbach(2012)Srebro, and Autschbach]{srebro_does_2012}
Srebro,~M.; Autschbach,~J. \emph{J. Phys. Chem. Lett.} \textbf{2012}, \emph{3},
  576--581\relax
\mciteBstWouldAddEndPuncttrue
\mciteSetBstMidEndSepPunct{\mcitedefaultmidpunct}
{\mcitedefaultendpunct}{\mcitedefaultseppunct}\relax
\EndOfBibitem
\bibitem[L\"uftner \latin{et~al.}(2014)L\"uftner, Refaely-Abramson, Pachler,
  Resel, Ramsey, Kronik, and Puschnig]{luftner_experimental_2014}
L\"uftner,~D.; Refaely-Abramson,~S.; Pachler,~M.; Resel,~R.; Ramsey,~M.~G.;
  Kronik,~L.; Puschnig,~P. \emph{Phys. Rev. B} \textbf{2014}, \emph{90},
  075204\relax
\mciteBstWouldAddEndPuncttrue
\mciteSetBstMidEndSepPunct{\mcitedefaultmidpunct}
{\mcitedefaultendpunct}{\mcitedefaultseppunct}\relax
\EndOfBibitem
\bibitem[Rohrdanz \latin{et~al.}(2009)Rohrdanz, Martins, and
  Herbert]{rohrdanz_long-range-corrected_2009}
Rohrdanz,~M.~A.; Martins,~K.~M.; Herbert,~J.~M. \emph{J. Chem. Phys.}
  \textbf{2009}, \emph{130}, 054112\relax
\mciteBstWouldAddEndPuncttrue
\mciteSetBstMidEndSepPunct{\mcitedefaultmidpunct}
{\mcitedefaultendpunct}{\mcitedefaultseppunct}\relax
\EndOfBibitem
\bibitem[Perdew \latin{et~al.}(1982)Perdew, Parr, Levy, and
  Balduz]{perdew_density-functional_1982}
Perdew,~J.~P.; Parr,~R.~G.; Levy,~M.; Balduz,~J.~L. \emph{Phys. Rev. Lett.}
  \textbf{1982}, \emph{49}, 1691--1694\relax
\mciteBstWouldAddEndPuncttrue
\mciteSetBstMidEndSepPunct{\mcitedefaultmidpunct}
{\mcitedefaultendpunct}{\mcitedefaultseppunct}\relax
\EndOfBibitem
\bibitem[Salzner and Baer(2009)Salzner, and Baer]{salzner_koopmans_2009}
Salzner,~U.; Baer,~R. \emph{J. Chem. Phys.} \textbf{2009}, \emph{131},
  231101\relax
\mciteBstWouldAddEndPuncttrue
\mciteSetBstMidEndSepPunct{\mcitedefaultmidpunct}
{\mcitedefaultendpunct}{\mcitedefaultseppunct}\relax
\EndOfBibitem
\bibitem[Levy \latin{et~al.}(1984)Levy, Perdew, and Sahni]{levy_exact_1984}
Levy,~M.; Perdew,~J.~P.; Sahni,~V. \emph{Phys. Rev. A} \textbf{1984},
  \emph{30}, 2745--2748\relax
\mciteBstWouldAddEndPuncttrue
\mciteSetBstMidEndSepPunct{\mcitedefaultmidpunct}
{\mcitedefaultendpunct}{\mcitedefaultseppunct}\relax
\EndOfBibitem
\bibitem[Perdew and Levy(1997)Perdew, and Levy]{perdew_comment_1997}
Perdew,~J.~P.; Levy,~M. \emph{Phys. Rev. B} \textbf{1997}, \emph{56},
  16021--16028\relax
\mciteBstWouldAddEndPuncttrue
\mciteSetBstMidEndSepPunct{\mcitedefaultmidpunct}
{\mcitedefaultendpunct}{\mcitedefaultseppunct}\relax
\EndOfBibitem
\bibitem[Almbladh and von Barth(1985)Almbladh, and von
  Barth]{almbladh_exact_1985}
Almbladh,~C.-O.; von Barth,~U. \emph{Phys. Rev. B} \textbf{1985}, \emph{31},
  3231--3244\relax
\mciteBstWouldAddEndPuncttrue
\mciteSetBstMidEndSepPunct{\mcitedefaultmidpunct}
{\mcitedefaultendpunct}{\mcitedefaultseppunct}\relax
\EndOfBibitem
\bibitem[Shao \latin{et~al.}(2015)Shao, Gan, Epifanovsky, Gilbert, Wormit,
  Kussmann, Lange, Behn, Deng, Feng, Ghosh, Goldey, Horn, Jacobson, Kaliman,
  Khaliullin, Ku\'s, Landau, Liu, Proynov, Rhee, Richard, Rohrdanz, Steele,
  Sundstrom, III, Zimmerman, Zuev, Albrecht, Alguire, Austin, Beran, Bernard,
  Berquist, Brandhorst, Bravaya, Brown, Casanova, Chang, Chen, Chien, Closser,
  Crittenden, Diedenhofen, Jr, Do, Dutoi, Edgar, Fatehi, Fusti-Molnar, Ghysels,
  Golubeva-Zadorozhnaya, Gomes, Hanson-Heine, Harbach, Hauser, Hohenstein,
  Holden, Jagau, Ji, Kaduk, Khistyaev, Kim, Kim, King, Klunzinger, Kosenkov,
  Kowalczyk, Krauter, Lao, Laurent, Lawler, Levchenko, Lin, Liu, Livshits,
  Lochan, Luenser, Manohar, Manzer, Mao, Mardirossian, Marenich, Maurer,
  Mayhall, Neuscamman, Oana, Olivares-Amaya, O’Neill, Parkhill, Perrine,
  Peverati, Prociuk, Rehn, Rosta, Russ, Sharada, Sharma, Small, Sodt, Stein,
  Stück, Su, Thom, Tsuchimochi, Vanovschi, Vogt, Vydrov, Wang, Watson, Wenzel,
  White, Williams, Yang, Yeganeh, Yost, You, Zhang, Zhang, Zhao, Brooks, Chan,
  Chipman, Cramer, III, Gordon, Hehre, Klamt, III, Schmidt, Sherrill, Truhlar,
  Warshel, Xu, Aspuru-Guzik, Baer, Bell, Besley, Chai, Dreuw, Dunietz, Furlani,
  Gwaltney, Hsu, Jung, Kong, Lambrecht, Liang, Ochsenfeld, Rassolov,
  Slipchenko, Subotnik, Voorhis, Herbert, Krylov, Gill, and
  Head-Gordon]{shao_advances_2015}
Shao,~Y.; Gan,~Z.; Epifanovsky,~E.; Gilbert,~A. T.~B.; Wormit,~M.;
  Kussmann,~J.; Lange,~A.~W.; Behn,~A.; Deng,~J.; Feng,~X.; Ghosh,~D.;
  Goldey,~M.; Horn,~P.~R.; Jacobson,~L.~D.; Kaliman,~I.; Khaliullin,~R.~Z.;
  Ku\'s,~T.; Landau,~A.; Liu,~J.; Proynov,~E.~I.; Rhee,~Y.~M.; Richard,~R.~M.;
  Rohrdanz,~M.~A.; Steele,~R.~P.; Sundstrom,~E.~J.; III,~H. L.~W.;
  Zimmerman,~P.~M.; Zuev,~D.; Albrecht,~B.; Alguire,~E.; Austin,~B.; Beran,~G.
  J.~O.; Bernard,~Y.~A.; Berquist,~E.; Brandhorst,~K.; Bravaya,~K.~B.;
  Brown,~S.~T.; Casanova,~D.; Chang,~C.-M.; Chen,~Y.; Chien,~S.~H.;
  Closser,~K.~D.; Crittenden,~D.~L.; Diedenhofen,~M.; Jr,~R. A.~D.; Do,~H.;
  Dutoi,~A.~D.; Edgar,~R.~G.; Fatehi,~S.; Fusti-Molnar,~L.; Ghysels,~A.;
  Golubeva-Zadorozhnaya,~A.; Gomes,~J.; Hanson-Heine,~M. W.~D.; Harbach,~P.
  H.~P.; Hauser,~A.~W.; Hohenstein,~E.~G.; Holden,~Z.~C.; Jagau,~T.-C.; Ji,~H.;
  Kaduk,~B.; Khistyaev,~K.; Kim,~J.; Kim,~J.; King,~R.~A.; Klunzinger,~P.;
  Kosenkov,~D.; Kowalczyk,~T.; Krauter,~C.~M.; Lao,~K.~U.; Laurent,~A.~D.;
  Lawler,~K.~V.; Levchenko,~S.~V.; Lin,~C.~Y.; Liu,~F.; Livshits,~E.;
  Lochan,~R.~C.; Luenser,~A.; Manohar,~P.; Manzer,~S.~F.; Mao,~S.-P.;
  Mardirossian,~N.; Marenich,~A.~V.; Maurer,~S.~A.; Mayhall,~N.~J.;
  Neuscamman,~E.; Oana,~C.~M.; Olivares-Amaya,~R.; O’Neill,~D.~P.;
  Parkhill,~J.~A.; Perrine,~T.~M.; Peverati,~R.; Prociuk,~A.; Rehn,~D.~R.;
  Rosta,~E.; Russ,~N.~J.; Sharada,~S.~M.; Sharma,~S.; Small,~D.~W.; Sodt,~A.;
  Stein,~T.; Stück,~D.; Su,~Y.-C.; Thom,~A. J.~W.; Tsuchimochi,~T.;
  Vanovschi,~V.; Vogt,~L.; Vydrov,~O.; Wang,~T.; Watson,~M.~A.; Wenzel,~J.;
  White,~A.; Williams,~C.~F.; Yang,~J.; Yeganeh,~S.; Yost,~S.~R.; You,~Z.-Q.;
  Zhang,~I.~Y.; Zhang,~X.; Zhao,~Y.; Brooks,~B.~R.; Chan,~G. K.~L.;
  Chipman,~D.~M.; Cramer,~C.~J.; III,~W. A.~G.; Gordon,~M.~S.; Hehre,~W.~J.;
  Klamt,~A.; III,~H. F.~S.; Schmidt,~M.~W.; Sherrill,~C.~D.; Truhlar,~D.~G.;
  Warshel,~A.; Xu,~X.; Aspuru-Guzik,~A.; Baer,~R.; Bell,~A.~T.; Besley,~N.~A.;
  Chai,~J.-D.; Dreuw,~A.; Dunietz,~B.~D.; Furlani,~T.~R.; Gwaltney,~S.~R.;
  Hsu,~C.-P.; Jung,~Y.; Kong,~J.; Lambrecht,~D.~S.; Liang,~W.; Ochsenfeld,~C.;
  Rassolov,~V.~A.; Slipchenko,~L.~V.; Subotnik,~J.~E.; Voorhis,~T.~V.;
  Herbert,~J.~M.; Krylov,~A.~I.; Gill,~P. M.~W.; Head-Gordon,~M. \emph{Mol.
  Phys.} \textbf{2015}, \emph{113}, 184\relax
\mciteBstWouldAddEndPuncttrue
\mciteSetBstMidEndSepPunct{\mcitedefaultmidpunct}
{\mcitedefaultendpunct}{\mcitedefaultseppunct}\relax
\EndOfBibitem
\bibitem[Becke(1993)]{becke_new_1993}
Becke,~A.~D. \emph{J. Chem. Phys.} \textbf{1993}, \emph{98}, 1372\relax
\mciteBstWouldAddEndPuncttrue
\mciteSetBstMidEndSepPunct{\mcitedefaultmidpunct}
{\mcitedefaultendpunct}{\mcitedefaultseppunct}\relax
\EndOfBibitem
\bibitem[Lee \latin{et~al.}(1988)Lee, Yang, and Parr]{lee_development_1988}
Lee,~C.; Yang,~W.; Parr,~R.~G. \emph{Phys. Rev. B} \textbf{1988}, \emph{37},
  785--789\relax
\mciteBstWouldAddEndPuncttrue
\mciteSetBstMidEndSepPunct{\mcitedefaultmidpunct}
{\mcitedefaultendpunct}{\mcitedefaultseppunct}\relax
\EndOfBibitem
\bibitem[Chelikowsky and Louie(1996)Chelikowsky, and
  Louie]{chelikowsky_quantum_1996}
Chelikowsky,~J.~R.; Louie,~S.~G. \emph{Quantum {Theory} {Real} {Mater.}};
  1996\relax
\mciteBstWouldAddEndPuncttrue
\mciteSetBstMidEndSepPunct{\mcitedefaultmidpunct}
{\mcitedefaultendpunct}{\mcitedefaultseppunct}\relax
\EndOfBibitem
\bibitem[Aryasetiawan and Gunnarsson(1998)Aryasetiawan, and
  Gunnarsson]{aryasetiawan_gw_1998}
Aryasetiawan,~F.; Gunnarsson,~O. \emph{Rep. Prog. Phys.} \textbf{1998},
  \emph{61}, 237\relax
\mciteBstWouldAddEndPuncttrue
\mciteSetBstMidEndSepPunct{\mcitedefaultmidpunct}
{\mcitedefaultendpunct}{\mcitedefaultseppunct}\relax
\EndOfBibitem
\bibitem[Onida \latin{et~al.}(2002)Onida, Reining, and
  Rubio]{onida_electronic_2002}
Onida,~G.; Reining,~L.; Rubio,~A. \emph{Rev. Mod. Phys.} \textbf{2002},
  \emph{74}, 601--659\relax
\mciteBstWouldAddEndPuncttrue
\mciteSetBstMidEndSepPunct{\mcitedefaultmidpunct}
{\mcitedefaultendpunct}{\mcitedefaultseppunct}\relax
\EndOfBibitem
\bibitem[Bruneval and Gatti(2014)Bruneval, and
  Gatti]{bruneval_quasiparticle_2014}
Bruneval,~F.; Gatti,~M. In \emph{First {Principles} {Approaches} to
  {Spectroscopic} {Properties} of {Complex} {Materials}}; Valentin,~C.~D.,
  Botti,~S., Cococcioni,~M., Eds.; Topics in {Current} {Chemistry} 347; 2014;
  pp 99--135\relax
\mciteBstWouldAddEndPuncttrue
\mciteSetBstMidEndSepPunct{\mcitedefaultmidpunct}
{\mcitedefaultendpunct}{\mcitedefaultseppunct}\relax
\EndOfBibitem
\bibitem[Hybertsen and Louie(1986)Hybertsen, and
  Louie]{hybertsen_electron_1986}
Hybertsen,~M.~S.; Louie,~S.~G. \emph{Phys. Rev. B} \textbf{1986}, \emph{34},
  5390\relax
\mciteBstWouldAddEndPuncttrue
\mciteSetBstMidEndSepPunct{\mcitedefaultmidpunct}
{\mcitedefaultendpunct}{\mcitedefaultseppunct}\relax
\EndOfBibitem
\bibitem[Bruneval and Marques(2013)Bruneval, and
  Marques]{bruneval_benchmarking_2013}
Bruneval,~F.; Marques,~M. A.~L. \emph{J. Chem. Theory Comput.} \textbf{2013},
  \emph{9}, 324\relax
\mciteBstWouldAddEndPuncttrue
\mciteSetBstMidEndSepPunct{\mcitedefaultmidpunct}
{\mcitedefaultendpunct}{\mcitedefaultseppunct}\relax
\EndOfBibitem
\bibitem[Atalla \latin{et~al.}(2013)Atalla, Yoon, Caruso, Rinke, and
  Scheffler]{atalla_hybrid_2013}
Atalla,~V.; Yoon,~M.; Caruso,~F.; Rinke,~P.; Scheffler,~M. \emph{Phys. Rev. B}
  \textbf{2013}, \emph{88}, 165122\relax
\mciteBstWouldAddEndPuncttrue
\mciteSetBstMidEndSepPunct{\mcitedefaultmidpunct}
{\mcitedefaultendpunct}{\mcitedefaultseppunct}\relax
\EndOfBibitem
\bibitem[K\"orbel \latin{et~al.}(2014)K\"orbel, Boulanger, Duchemin, Blase,
  Marques, and Botti]{korbel_benchmark_2014}
K\"orbel,~S.; Boulanger,~P.; Duchemin,~I.; Blase,~X.; Marques,~M. A.~L.;
  Botti,~S. \emph{J. Chem. Theory Comput.} \textbf{2014}, \emph{10},
  3934--3943\relax
\mciteBstWouldAddEndPuncttrue
\mciteSetBstMidEndSepPunct{\mcitedefaultmidpunct}
{\mcitedefaultendpunct}{\mcitedefaultseppunct}\relax
\EndOfBibitem
\bibitem[Govoni and Galli(2015)Govoni, and Galli]{govoni_large_2015}
Govoni,~M.; Galli,~G. \emph{J. Chem. Theory Comput.} \textbf{2015}, \emph{11},
  2680--2696\relax
\mciteBstWouldAddEndPuncttrue
\mciteSetBstMidEndSepPunct{\mcitedefaultmidpunct}
{\mcitedefaultendpunct}{\mcitedefaultseppunct}\relax
\EndOfBibitem
\bibitem[Bruneval \latin{et~al.}(2015)Bruneval, Hamed, and
  Neaton]{bruneval_systematic_2015}
Bruneval,~F.; Hamed,~S.~M.; Neaton,~J.~B. \emph{J. Chem. Phys.} \textbf{2015},
  \emph{142}, 244101\relax
\mciteBstWouldAddEndPuncttrue
\mciteSetBstMidEndSepPunct{\mcitedefaultmidpunct}
{\mcitedefaultendpunct}{\mcitedefaultseppunct}\relax
\EndOfBibitem
\bibitem[Perdew \latin{et~al.}(1996)Perdew, Ernzerhof, and
  Burke]{perdew_rationale_1996}
Perdew,~J.~P.; Ernzerhof,~M.; Burke,~K. \emph{J. Chem. Phys.} \textbf{1996},
  \emph{105}, 9982\relax
\mciteBstWouldAddEndPuncttrue
\mciteSetBstMidEndSepPunct{\mcitedefaultmidpunct}
{\mcitedefaultendpunct}{\mcitedefaultseppunct}\relax
\EndOfBibitem
\bibitem[Stan \latin{et~al.}(2006)Stan, Dahlen, and {van
  Leeuwen}]{stan_fully_2006}
Stan,~A.; Dahlen,~N.~E.; {van Leeuwen},~R. \emph{Europhys. Lett.}
  \textbf{2006}, \emph{76}, 298--304\relax
\mciteBstWouldAddEndPuncttrue
\mciteSetBstMidEndSepPunct{\mcitedefaultmidpunct}
{\mcitedefaultendpunct}{\mcitedefaultseppunct}\relax
\EndOfBibitem
\bibitem[Rostgaard \latin{et~al.}(2010)Rostgaard, Jacobsen, and
  Thygesen]{rostgaard_fully_2010}
Rostgaard,~C.; Jacobsen,~K.~W.; Thygesen,~K.~S. \emph{Phys. Rev. B}
  \textbf{2010}, \emph{81}, 085103\relax
\mciteBstWouldAddEndPuncttrue
\mciteSetBstMidEndSepPunct{\mcitedefaultmidpunct}
{\mcitedefaultendpunct}{\mcitedefaultseppunct}\relax
\EndOfBibitem
\bibitem[Caruso \latin{et~al.}(2012)Caruso, Rinke, Ren, Scheffler, and
  Rubio]{caruso_unified_2012}
Caruso,~F.; Rinke,~P.; Ren,~X.; Scheffler,~M.; Rubio,~A. \emph{Phys. Rev. B}
  \textbf{2012}, \emph{86}, 081102\relax
\mciteBstWouldAddEndPuncttrue
\mciteSetBstMidEndSepPunct{\mcitedefaultmidpunct}
{\mcitedefaultendpunct}{\mcitedefaultseppunct}\relax
\EndOfBibitem
\bibitem[Jacquemin \latin{et~al.}(2015)Jacquemin, Duchemin, and
  Blase]{jacquemin_benchmarking_2015}
Jacquemin,~D.; Duchemin,~I.; Blase,~X. \emph{J. Chem. Theory Comput.}
  \textbf{2015}, \emph{11}, 3290\relax
\mciteBstWouldAddEndPuncttrue
\mciteSetBstMidEndSepPunct{\mcitedefaultmidpunct}
{\mcitedefaultendpunct}{\mcitedefaultseppunct}\relax
\EndOfBibitem
\bibitem[Shishkin and Kresse(2007)Shishkin, and Kresse]{shishkin_kresse_2007}
Shishkin,~M.; Kresse,~G. \emph{Phys. Rev. B} \textbf{2007}, \emph{75},
  235102\relax
\mciteBstWouldAddEndPuncttrue
\mciteSetBstMidEndSepPunct{\mcitedefaultmidpunct}
{\mcitedefaultendpunct}{\mcitedefaultseppunct}\relax
\EndOfBibitem
\bibitem[Bruneval(2015)]{molgw_cite}
Bruneval,~F. MOLGW: A slow but accurate many-body perturbation theory code.
  2015; \url{https://github.com/bruneval/molgw}, (Visited on 03/09/2016).\relax
\mciteBstWouldAddEndPunctfalse
\mciteSetBstMidEndSepPunct{\mcitedefaultmidpunct}
{}{\mcitedefaultseppunct}\relax
\EndOfBibitem
\bibitem[Bruneval(2012)]{bruneval_ionization_2012}
Bruneval,~F. \emph{J. Chem. Phys.} \textbf{2012}, \emph{136}, 194107\relax
\mciteBstWouldAddEndPuncttrue
\mciteSetBstMidEndSepPunct{\mcitedefaultmidpunct}
{\mcitedefaultendpunct}{\mcitedefaultseppunct}\relax
\EndOfBibitem
\bibitem[Valeev(2015)]{libint2}
Valeev,~E.~F. A library for the evaluation of molecular integrals of many-body
  operators over Gaussian functions. 2015; \url{http://libint.valeyev.net/},
  (Visited on 03/09/2016).\relax
\mciteBstWouldAddEndPunctfalse
\mciteSetBstMidEndSepPunct{\mcitedefaultmidpunct}
{}{\mcitedefaultseppunct}\relax
\EndOfBibitem
\bibitem[Marques \latin{et~al.}(2012)Marques, Oliveira, and
  Burnus]{marques_libxc:_2012}
Marques,~M. A.~L.; Oliveira,~M. J.~T.; Burnus,~T. \emph{Comput. Phys. Commun.}
  \textbf{2012}, \emph{183}, 2272\relax
\mciteBstWouldAddEndPuncttrue
\mciteSetBstMidEndSepPunct{\mcitedefaultmidpunct}
{\mcitedefaultendpunct}{\mcitedefaultseppunct}\relax
\EndOfBibitem
\bibitem[Vahtras \latin{et~al.}(1993)Vahtras, Alml{\"o}f, and
  Feyereisen]{vahtras_cpl1993}
Vahtras,~O.; Alml{\"o}f,~J.; Feyereisen,~M. \emph{Chem. Phys. Lett.}
  \textbf{1993}, \emph{213}, 514 -- 518\relax
\mciteBstWouldAddEndPuncttrue
\mciteSetBstMidEndSepPunct{\mcitedefaultmidpunct}
{\mcitedefaultendpunct}{\mcitedefaultseppunct}\relax
\EndOfBibitem
\bibitem[Weigend(2002)]{weigend_pccp2002}
Weigend,~F. \emph{Phys. Chem. Chem. Phys.} \textbf{2002}, \emph{4},
  4285--4291\relax
\mciteBstWouldAddEndPuncttrue
\mciteSetBstMidEndSepPunct{\mcitedefaultmidpunct}
{\mcitedefaultendpunct}{\mcitedefaultseppunct}\relax
\EndOfBibitem
\bibitem[Rohlfing \latin{et~al.}(1995)Rohlfing, Kr\"uger, and
  Pollmann]{rohlfing_prb1995}
Rohlfing,~M.; Kr\"uger,~P.; Pollmann,~J. \emph{Phys. Rev. B} \textbf{1995},
  \emph{52}, 1905--1917\relax
\mciteBstWouldAddEndPuncttrue
\mciteSetBstMidEndSepPunct{\mcitedefaultmidpunct}
{\mcitedefaultendpunct}{\mcitedefaultseppunct}\relax
\EndOfBibitem
\bibitem[Ren \latin{et~al.}(2012)Ren, Rinke, Blum, Wieferink, Tkatchenko,
  Sanfilippo, Reuter, and Scheffler]{ren_njp2012}
Ren,~X.; Rinke,~P.; Blum,~V.; Wieferink,~J.; Tkatchenko,~A.; Sanfilippo,~A.;
  Reuter,~K.; Scheffler,~M. \emph{New J. Phys.} \textbf{2012}, \emph{14},
  053020\relax
\mciteBstWouldAddEndPuncttrue
\mciteSetBstMidEndSepPunct{\mcitedefaultmidpunct}
{\mcitedefaultendpunct}{\mcitedefaultseppunct}\relax
\EndOfBibitem
\bibitem[Bruneval \latin{et~al.}()Bruneval, Rangel, Hamed, Shao, Yang, and
  Neaton]{molgw1}
Bruneval,~F.; Rangel,~T.; Hamed,~S.~M.; Shao,~M.; Yang,~C.; Neaton,~J.~B. in
  preparation.\relax
\mciteBstWouldAddEndPunctfalse
\mciteSetBstMidEndSepPunct{\mcitedefaultmidpunct}
{}{\mcitedefaultseppunct}\relax
\EndOfBibitem
\bibitem[Dunning(1989)]{dunning_jcp1989}
Dunning,~T.~H. \emph{J. Chem. Phys.} \textbf{1989}, \emph{90}, 1007--1023\relax
\mciteBstWouldAddEndPuncttrue
\mciteSetBstMidEndSepPunct{\mcitedefaultmidpunct}
{\mcitedefaultendpunct}{\mcitedefaultseppunct}\relax
\EndOfBibitem
\bibitem[Frisch \latin{et~al.}()Frisch, Trucks, Schlegel, Scuseria, Robb,
  Cheeseman, Scalmani, Barone, Mennucci, Petersson, Nakatsuji, Caricato, Li,
  Hratchian, Izmaylov, Bloino, Zheng, Sonnenberg, Hada, Ehara, Toyota, Fukuda,
  Hasegawa, Ishida, Nakajima, Honda, Kitao, Nakai, Vreven, Montgomery, Peralta,
  Ogliaro, Bearpark, Heyd, Brothers, Kudin, Staroverov, Kobayashi, Normand,
  Raghavachari, Rendell, Burant, Iyengar, Tomasi, Cossi, Rega, Millam, Klene,
  Knox, Cross, Bakken, Adamo, Jaramillo, Gomperts, Stratmann, Yazyev, Austin,
  Cammi, Pomelli, Ochterski, Martin, Morokuma, Zakrzewski, Voth, Salvador,
  Dannenberg, Dapprich, Daniels, Farkas, Foresman, Ortiz, Cioslowski, and
  Fox]{g09}
Frisch,~M.~J.; Trucks,~G.~W.; Schlegel,~H.~B.; Scuseria,~G.~E.; Robb,~M.~A.;
  Cheeseman,~J.~R.; Scalmani,~G.; Barone,~V.; Mennucci,~B.; Petersson,~G.~A.;
  Nakatsuji,~H.; Caricato,~M.; Li,~X.; Hratchian,~H.~P.; Izmaylov,~A.~F.;
  Bloino,~J.; Zheng,~G.; Sonnenberg,~J.~L.; Hada,~M.; Ehara,~M.; Toyota,~K.;
  Fukuda,~R.; Hasegawa,~J.; Ishida,~M.; Nakajima,~T.; Honda,~Y.; Kitao,~O.;
  Nakai,~H.; Vreven,~T.; Montgomery,~J.~A.,~{Jr.}; Peralta,~J.~E.; Ogliaro,~F.;
  Bearpark,~M.; Heyd,~J.~J.; Brothers,~E.; Kudin,~K.~N.; Staroverov,~V.~N.;
  Kobayashi,~R.; Normand,~J.; Raghavachari,~K.; Rendell,~A.; Burant,~J.~C.;
  Iyengar,~S.~S.; Tomasi,~J.; Cossi,~M.; Rega,~N.; Millam,~J.~M.; Klene,~M.;
  Knox,~J.~E.; Cross,~J.~B.; Bakken,~V.; Adamo,~C.; Jaramillo,~J.;
  Gomperts,~R.; Stratmann,~R.~E.; Yazyev,~O.; Austin,~A.~J.; Cammi,~R.;
  Pomelli,~C.; Ochterski,~J.~W.; Martin,~R.~L.; Morokuma,~K.;
  Zakrzewski,~V.~G.; Voth,~G.~A.; Salvador,~P.; Dannenberg,~J.~J.;
  Dapprich,~S.; Daniels,~A.~D.; Farkas,~{\"O}.; Foresman,~J.~B.; Ortiz,~J.~V.;
  Cioslowski,~J.; Fox,~D.~J. Gaussian∼09 {R}evision {E}.01. Gaussian Inc.
  Wallingford CT 2009\relax
\mciteBstWouldAddEndPuncttrue
\mciteSetBstMidEndSepPunct{\mcitedefaultmidpunct}
{\mcitedefaultendpunct}{\mcitedefaultseppunct}\relax
\EndOfBibitem
\bibitem[Hajgat\'o \latin{et~al.}(2009)Hajgat\'o, Szieberth, Geerlings, Prof~t,
  and Deleuze]{hajgato_benchmark_2009}
Hajgat\'o,~B.; Szieberth,~D.; Geerlings,~P.; Prof~t,~F.~D.; Deleuze,~M.~S.
  \emph{J. Chem. Phys.} \textbf{2009}, \emph{131}, 224321\relax
\mciteBstWouldAddEndPuncttrue
\mciteSetBstMidEndSepPunct{\mcitedefaultmidpunct}
{\mcitedefaultendpunct}{\mcitedefaultseppunct}\relax
\EndOfBibitem
\bibitem[Frank \latin{et~al.}(1988)Frank, Yannoulis, Dudde, and
  Koch]{frank_unoccupied_1988}
Frank,~K.~H.; Yannoulis,~P.; Dudde,~R.; Koch,~E.~E. \emph{J. Chem. Phys.}
  \textbf{1988}, \emph{89}, 7569\relax
\mciteBstWouldAddEndPuncttrue
\mciteSetBstMidEndSepPunct{\mcitedefaultmidpunct}
{\mcitedefaultendpunct}{\mcitedefaultseppunct}\relax
\EndOfBibitem
\bibitem[Pope and Swenberg(1999)Pope, and Swenberg]{pope_electronic_1999}
Pope,~M.; Swenberg,~C.~E. \emph{Electronic {Processes} in {Organic} {Crystals}
  and {Polymer s}}, 2nd ed.; Oxford University Press: New York, 1999\relax
\mciteBstWouldAddEndPuncttrue
\mciteSetBstMidEndSepPunct{\mcitedefaultmidpunct}
{\mcitedefaultendpunct}{\mcitedefaultseppunct}\relax
\EndOfBibitem
\bibitem[nis()]{nist}
Standard Reference Data. NIST. \url{http://www.nist.gov/}, (Visited on
  03/09/2016).\relax
\mciteBstWouldAddEndPunctfalse
\mciteSetBstMidEndSepPunct{\mcitedefaultmidpunct}
{}{\mcitedefaultseppunct}\relax
\EndOfBibitem
\bibitem[Heinis \latin{et~al.}(1993)Heinis, Chowdhury, and
  Kebarle]{heinis_electron_1993}
Heinis,~T.; Chowdhury,~S.; Kebarle,~P. \emph{Org. Mass Spectrom.}
  \textbf{1993}, \emph{28}, 358--365\relax
\mciteBstWouldAddEndPuncttrue
\mciteSetBstMidEndSepPunct{\mcitedefaultmidpunct}
{\mcitedefaultendpunct}{\mcitedefaultseppunct}\relax
\EndOfBibitem
\bibitem[Whittleton \latin{et~al.}(2015)Whittleton, Vazquez, Isborn, and
  Johnson]{whittleton_density-functional_2015}
Whittleton,~S.~R.; Vazquez,~X. A.~S.; Isborn,~C.~M.; Johnson,~E.~R. \emph{J.
  Chem. Phys.} \textbf{2015}, \emph{142}, 184106\relax
\mciteBstWouldAddEndPuncttrue
\mciteSetBstMidEndSepPunct{\mcitedefaultmidpunct}
{\mcitedefaultendpunct}{\mcitedefaultseppunct}\relax
\EndOfBibitem
\bibitem[{Mori-S\'anchez} \latin{et~al.}(2008){Mori-S\'anchez}, Cohen, and
  Yang]{mori-sanchez_localization_2008}
{Mori-S\'anchez},~P.; Cohen,~A.~J.; Yang,~W. \emph{Phys. Rev. Lett.}
  \textbf{2008}, \emph{100}, 146401\relax
\mciteBstWouldAddEndPuncttrue
\mciteSetBstMidEndSepPunct{\mcitedefaultmidpunct}
{\mcitedefaultendpunct}{\mcitedefaultseppunct}\relax
\EndOfBibitem
\bibitem[Marsili \latin{et~al.}(2013)Marsili, Botti, Palummo, Degoli, Pulci,
  Weissker, Marques, Ossicini, and Del~Sole]{marsili_ab_2013}
Marsili,~M.; Botti,~S.; Palummo,~M.; Degoli,~E.; Pulci,~O.; Weissker,~H.-C.;
  Marques,~M. A.~L.; Ossicini,~S.; Del~Sole,~R. \emph{J. Phys. Chem. C}
  \textbf{2013}, 14229\relax
\mciteBstWouldAddEndPuncttrue
\mciteSetBstMidEndSepPunct{\mcitedefaultmidpunct}
{\mcitedefaultendpunct}{\mcitedefaultseppunct}\relax
\EndOfBibitem
\bibitem[Gavnholt \latin{et~al.}(2008)Gavnholt, Olsen, Engelund, and
  Schi{\o}tz]{gavnholt_delta_2008}
Gavnholt,~J.; Olsen,~T.; Engelund,~M.; Schi{\o}tz,~J. \emph{Phys. Rev. B}
  \textbf{2008}, \emph{78}, 075441\relax
\mciteBstWouldAddEndPuncttrue
\mciteSetBstMidEndSepPunct{\mcitedefaultmidpunct}
{\mcitedefaultendpunct}{\mcitedefaultseppunct}\relax
\EndOfBibitem
\bibitem[Stein \latin{et~al.}(2010)Stein, Eisenberg, Kronik, and
  Baer]{stein_fundamental_2010}
Stein,~T.; Eisenberg,~H.; Kronik,~L.; Baer,~R. \emph{Phys. Rev. Lett.}
  \textbf{2010}, \emph{105}, 266802\relax
\mciteBstWouldAddEndPuncttrue
\mciteSetBstMidEndSepPunct{\mcitedefaultmidpunct}
{\mcitedefaultendpunct}{\mcitedefaultseppunct}\relax
\EndOfBibitem
\bibitem[van Setten \latin{et~al.}(2015)van Setten, Caruso, Sharifzadeh, Ren,
  Scheffler, Liu, Lischner, Lin, Deslippe, Louie, Yang, Weigend, Neaton, Evers,
  and Rinke]{van_setten_gw100:_2015}
van Setten,~M.~J.; Caruso,~F.; Sharifzadeh,~S.; Ren,~X.; Scheffler,~M.;
  Liu,~F.; Lischner,~J.; Lin,~L.; Deslippe,~J.~R.; Louie,~S.~G.; Yang,~C.;
  Weigend,~F.; Neaton,~J.~B.; Evers,~F.; Rinke,~P. \emph{J. Chem. Theory
  Comput.} \textbf{2015}, \emph{11}, 5665--5687\relax
\mciteBstWouldAddEndPuncttrue
\mciteSetBstMidEndSepPunct{\mcitedefaultmidpunct}
{\mcitedefaultendpunct}{\mcitedefaultseppunct}\relax
\EndOfBibitem
\bibitem[Sharifzadeh \latin{et~al.}(2012)Sharifzadeh, Tamblyn, Doak, Darancet,
  and Neaton]{sharifzadeh_quantitative_2012}
Sharifzadeh,~S.; Tamblyn,~I.; Doak,~P.; Darancet,~P.~T.; Neaton,~J.~B.
  \emph{Eur. Phys. J. B} \textbf{2012}, \emph{85}, 1\relax
\mciteBstWouldAddEndPuncttrue
\mciteSetBstMidEndSepPunct{\mcitedefaultmidpunct}
{\mcitedefaultendpunct}{\mcitedefaultseppunct}\relax
\EndOfBibitem
\bibitem[Lischner \latin{et~al.}(2014)Lischner, Sharifzadeh, Deslippe, Neaton,
  and Louie]{lischner_effects_2014}
Lischner,~J.; Sharifzadeh,~S.; Deslippe,~J.~c.; Neaton,~J.~B.; Louie,~S.~G.
  \emph{Phys. Rev. B} \textbf{2014}, \emph{90}, 115130\relax
\mciteBstWouldAddEndPuncttrue
\mciteSetBstMidEndSepPunct{\mcitedefaultmidpunct}
{\mcitedefaultendpunct}{\mcitedefaultseppunct}\relax
\EndOfBibitem
\bibitem[Bendikov \latin{et~al.}(2004)Bendikov, Duong, Starkey, Houk, Carter,
  and Wudl]{bendikov_oligoacenes:_2004}
Bendikov,~M.; Duong,~H.~M.; Starkey,~K.; Houk,~K.~N.; Carter,~E.~A.; Wudl,~F.
  \emph{J. Am. Chem. Soc.} \textbf{2004}, \emph{126}, 7416--7417\relax
\mciteBstWouldAddEndPuncttrue
\mciteSetBstMidEndSepPunct{\mcitedefaultmidpunct}
{\mcitedefaultendpunct}{\mcitedefaultseppunct}\relax
\EndOfBibitem
\bibitem[Caruso \latin{et~al.}(2013)Caruso, Rohr, Hellgren, Ren, Rinke, Rubio,
  and Scheffler]{caruso_bond_2013}
Caruso,~F.; Rohr,~D.~R.; Hellgren,~M.; Ren,~X.; Rinke,~P.; Rubio,~A.;
  Scheffler,~M. \emph{Phys. Rev. Lett.} \textbf{2013}, \emph{110}, 146403\relax
\mciteBstWouldAddEndPuncttrue
\mciteSetBstMidEndSepPunct{\mcitedefaultmidpunct}
{\mcitedefaultendpunct}{\mcitedefaultseppunct}\relax
\EndOfBibitem
\end{mcitethebibliography}

\begin{tocentry}
\centering
\includegraphics{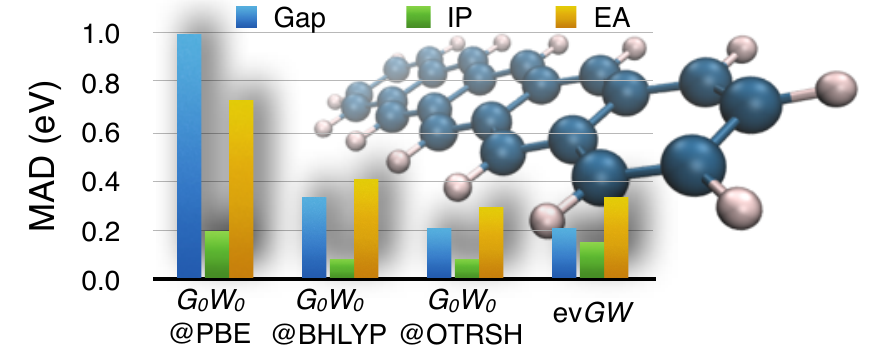}
\end{tocentry}

\end{document}